\pdfoutput=1
\newcommand*{\ATLASLATEXPATH}{}
\documentclass[USenglish,texlive=2016,cernpreprint]{\ATLASLATEXPATH atlasdoc}
\usepackage[subfigure=true]{\ATLASLATEXPATH atlaspackage}
\usepackage{\ATLASLATEXPATH atlasbiblatex}
\usepackage{\ATLASLATEXPATH atlasphysics}

\usepackage{EXOT-2017-11-PAPER-defs}

\usepackage{longtable}
\usepackage{comment}
\usepackage{siunitx}
\usepackage{xspace}
\usepackage{booktabs}
\PassOptionsToPackage{hyphens}{url}
\usepackage{hyperref}

\AtlasTitle{Search for four-top-quark production in the single-lepton and opposite-sign dilepton final states in $pp$ collisions at $\sqrt{s} = 13 \TeV$ with the ATLAS detector}
\author{The ATLAS Collaboration}
\AtlasRefCode{EXOT-2017-11}
\PreprintIdNumber{CERN-EP-2018-174}
\AtlasJournalRef{Phys. Rev. D 99, 052009 (2019)}
\AtlasDOI{10.1103/PhysRevD.99.052009}

\AtlasAbstract{%
A search for four-top-quark production, \tttt, is presented. It is based on proton--proton collision data with a center-of-mass energy $\sqrt{s} = 13\TeV$ collected by the ATLAS detector at the Large Hadron Collider during the years 2015 and 2016, corresponding to an integrated luminosity of \totlumi. Data are analyzed in both the single-lepton and opposite-sign dilepton channels, characterized by the presence of one or two isolated electrons or muons with high-transverse momentum and multiple jets. A data-driven method is used to estimate the dominant background from top-quark pair production in association with jets. No significant excess above the Standard Model expectation is observed. The result is combined with the previous same-sign dilepton and multilepton searches carried out by the ATLAS Collaboration and an observed (expected) upper limit of $5.3$ ($2.1$) times the four-top-quark Standard Model cross section is obtained at 95\% confidence level. Additionally, an upper limit on the anomalous four-top-quark production cross section is set in the context of an effective field theory model.}

\hypersetup{pdftitle={ATLAS document},pdfauthor={The ATLAS Collaboration}}

\begin{document}

\maketitle

\section{Introduction}
With a mass close to the scale of electroweak symmetry breaking, the top quark, besides having a large coupling to the Standard Model (SM) Higgs boson~\cite{cms_ttH_obs,atlas_ttH_obs}, is predicted to have large couplings to new particles hypothesized in many models beyond the Standard Model (BSM)~\cite{Susskind:1978ms,Kaplan:1983sm,ArkaniHamed:2002qy}. Possible new phenomena may enhance the cross sections over SM predictions for various processes involving top quarks, and in particular for the production of four top quarks~\cite{guchait:2007jd,pomarol:2008bh,plehn:2008ae,Kilic:2009mi,degrande:2010kt,cacciapaglia:2011kz,DeSimone:2012fs,Deandrea:2014raa,Alvarez,cacciapaglia:2009pa}. This paper focuses on a search for four-top-quark (\tttt) production via the SM processes in proton--proton ($pp$) collisions at the Large Hadron Collider (LHC) and the results are interpreted in the context of an effective field theory (EFT) approach where the BSM contribution is represented via a four-top-quarks contact interaction~\cite{degrande:2010kt}.

The SM four-top-quark production cross section ($\sigma_{\mathrm{SM}}^{\tttt}$) in $pp$ collisions at a center-of-mass energy $\sqrt{s} = 13 $ TeV is predicted to be $\sigma_{\mathrm{SM}}^{\tttt} =  9.2 $ fb at next-to-leading-order (NLO) accuracy in QCD, with scale and parton distribution function (PDF) uncertainties of the order of 30\% and 6\%, respectively~\cite{bevilacqua:2012constraining, Alwall:2014hca}. Previous searches for four-top-quark production using LHC Run~2 data at $\sqrt{s} = 13 $ TeV were performed by both the ATLAS~\cite{Aaboud:2017faq,HTX,SS_4tops} and CMS~\cite{Khachatryan:2016kod,Sirunyan:2017tep,Sirunyan:2017uyt,Sirunyan:2017roi} Collaborations. Among them, the most sensitive one is a CMS search~\cite{Sirunyan:2017roi} obtaining an observed (expected) 95\% confidence level (C.L.) upper limit of 4.5 (2.3) times the SM expectation. Searches for anomalous \tttt\ production via an EFT model were recently performed by the ATLAS Collaboration~\cite{HTX,SS_4tops}, which set an observed (expected) upper limit of 16~fb (31~fb) on the production cross section at 95\% C.L.

The four-top-quark production events can give rise to different final states depending on the hadronic or semileptonic decay mode of each of the top quarks. The four-top-quark decay topology considered in this search corresponds to either single-lepton events with one isolated charged lepton (electron or muon)\footnote{The $\tau$-leptons are not reconstructed; however, isolated electron and muon by-products of the leptonically decaying $\tau$-leptons are considered in the analysis.} or dilepton events with two opposite-sign charged leptons (electrons or muons). The event topology also features high jet multiplicity and high multiplicity of jets containing $b$-hadrons. Signal events are characterized by high scalar sum of the jet transverse momenta ($\hthad$), which provides good discrimination against the dominant background, i.e.~top-quark pair production in association with jets ($\ttbar$+jets). Given that the four-top-quark production events are expected to contain hadronically decaying top-quark candidates with collimated or partially collimated topologies, the analysis makes use of ``mass-tagged reclustered large-$R$ (RCLR) jets''~\cite{Nachman:2014kla,ATLAS-CONF-2017-062}, which will be described in Section~\ref{sec:ObjReco}.

Selected events in each of the two channels are classified into several categories according to the number of jets, $b$-tagged jets and \mreclust\ jets. A data-driven method is developed to estimate the dominant $\ttbar$+jets background. The Monte Carlo (MC) simulation is used in order to estimate correction factors and evaluate the systematic uncertainties of the data-driven estimate.

The paper is organized as follows: the ATLAS detector is described in Section~\ref{section:ATLASDet}. Section~\ref{sec:ObjReco} summarizes the selection criteria applied to events and reconstructed objects. The simulation-based signal and background modeling, together with the data-driven estimation of non-prompt and fake lepton backgrounds are discussed in Section~\ref{sec:SandB}. Section~\ref{sec:AnalysisMethod} is devoted to the search strategy and classification of event topologies, while the $\ttbar$+jets background estimation technique using data is described in Section~\ref{sec:ttDD}. The systematic uncertainties are summarized in Section~\ref{sec:Systematics}. Section~\ref{sec:Results} presents the results and the combination with the same-sign dilepton and multilepton final-states search~\cite{SS_4tops} carried out by ATLAS.

\section{ATLAS detector}
\label{section:ATLASDet}
The ATLAS detector~\cite{Aad:2008zzm} at the LHC is a multipurpose particle detector with a forward--backward symmetric cylindrical geometry and nearly 4$\pi$ coverage in solid angle.\footnote{ATLAS uses a right-handed coordinate system with its origin at the nominal interaction point (IP) in the center of the detector. The positive $x$-axis is defined by the direction from the IP to the center of the LHC ring, with the positive $y$-axis pointing upwards, while the beam direction defines the $z$-axis. Cylindrical coordinates $(r,\phi)$ are used in the transverse plane, $\phi$ being the azimuthal angle around the $z$-axis. The pseudorapidity $\eta$ is defined in terms of the polar angle $\theta$ by $\eta=-\ln\tan(\theta/2)$. Unless stated otherwise, the angular distance is defined as $\Delta R \equiv \sqrt{ (\Delta \eta)^2 + (\Delta \phi)^2} $.} It consists of an inner tracking detector (ID), electromagnetic and hadronic calorimeters, and a muon spectrometer. The inner detector, including the insertable B-layer~\cite{Capeans:1291633,IBL}, provides charged-particle tracking from silicon pixel and microstrip detectors in the pseudorapidity region $|\eta|<2.5$, surrounded by a transition radiation tracker that enhances electron identification in the region $|\eta|<2.0$. The ID is surrounded by a thin superconducting solenoid providing an axial 2 T magnetic field, and by a fine-granularity lead/liquid-argon electromagnetic calorimeter covering $|\eta|<3.2$, which provides energy measurements of electromagnetic showers. Hadron calorimetry is also based on the sampling technique and covers $|\eta|<4.9$, with either scintillator tiles or liquid argon as the active medium and with steel, copper or tungsten as the absorber material. An extensive muon spectrometer with an air-core toroid magnet system surrounds the calorimeters. It includes three layers of high-precision tracking chambers, which provide coverage in the range $|\eta|<2.7$. The field integral of the toroid magnets ranges from 2.0 to 6.0 Tm across most of the detector. A two-level trigger system~\cite{Aaboud:2016leb}, the first level using custom hardware and followed by a software-based level, is used to reduce the event rate to a maximum of around one kHz for offline storage.

\section{Object and event selection}
\label{sec:ObjReco}
Events are selected from $pp$ collisions with $\sqrt{s} = 13$ TeV recorded by the ATLAS detector in 2015 and 2016. Only events for which all relevant subsystems were operational are considered. The data set corresponds to an integrated luminosity of $36.1 \pm 0.8$\ fb$^{-1}$~\cite{Aaboud:2016hhf}. The event reconstruction is affected by multiple inelastic $pp$ collisions in a single bunch crossing and by collisions in neighboring bunch crossings, referred to as ``pileup''. The number of interactions per bunch crossing in this data set ranges from about 8 to 45 interactions. Events are required to have at least one reconstructed vertex with two or more associated tracks with transverse momentum $\pt > 0.4 \gev$. If multiple vertices are reconstructed, the vertex with the largest sum of the squares of the transverse momenta of associated tracks is taken as the primary vertex~\cite{ref-pv}.

Events in both the single-lepton and dilepton channels were recorded using single-lepton triggers. Events were selected using triggers with either low $\pt$ thresholds and a lepton-isolation requirement, or with higher thresholds but with a looser identification criterion and without any isolation requirement. The lowest $\pt$ threshold used for muons is $20$ (26) GeV in 2015 (2016), while the higher $\pt$ threshold is $50\gev$ in both years. For electrons, triggers with a $\pt$ threshold of 24 (26) GeV in 2015 (2016) and isolation requirements are used along with triggers with a $60\gev$ threshold and no isolation requirement, and with a 120 (140)$\gev$ threshold with looser identification criteria.

Electron candidates are reconstructed~\cite{Aad:2014nim,ATLAS-CONF-2016-024} from an isolated electromagnetic calorimeter energy deposit, matched to a track in the ID, within the fiducial region of $ | \eta_{\mathrm{cluster}} | < 2.47 $, where $\eta_{\mathrm{cluster}}$ is the pseudorapidity of the calorimeter energy deposit associated with the electron candidate. Candidates within the transition region between the barrel and endcap electromagnetic calorimeters, $ 1.37 < | \eta_{\mathrm{cluster}} | < 1.52 $, are excluded. The electron candidates are required to have $\pt > 30 \gev$ and to satisfy ``tight'' likelihood-based identification criteria~\cite{ATLAS-CONF-2016-024} based on calorimeter, tracking and combined variables that provide good separation between electrons and jets. 
Muon candidates are reconstructed~\cite{muon_latest} by combining tracks reconstructed in both the ID and the muon spectrometer. Candidates are required to pass the ``medium'' identification criteria~\cite{muon_latest} and to have $\pt > 30 \gev$ and $|\eta| < 2.5$. To reduce the contribution from non-prompt leptons (e.g. from semileptonic $b$- or $c$-hadron decays), photon conversions and hadrons, lepton candidates are also required to be isolated. The lepton isolation is estimated using the scalar sum of all tracks excluding the lepton candidate itself ($I_R = \sum \pt^{\textrm{trk}}$) within a cone defined by $\Delta R<R_{\textrm{cut}}$ along the direction of the lepton. The value of $R_{\textrm{cut}}$ is the smaller of $r_{\textrm{min}}$ and $10\gev/\pt^\ell$, where $r_{\textrm{min}}$ is set to 0.2 (0.3) for electron (muon) candidates, and $\pt^\ell$ is the lepton $\pt$. All lepton candidates are required to satisfy $I_R/\pt^\ell < 0.06$. Finally, lepton tracks must match the primary vertex of the event: the longitudinal impact parameter $z_{0}$ is required to satisfy $|z_{0}\sin\theta|< 0.5$~mm, where $\theta$ is the polar angle of the track. The transverse impact parameter significance $|d_{0}| / \sigma(d_{0})$ must be less than 5 for electrons and 3 for muons, where $d_0$ is the transverse impact parameter and $\sigma(d_0)$ is its uncertainty.

Jet candidates are reconstructed from three-dimensional topological energy clusters~\cite{Aad:2016upy} in the calorimeter using the anti-$k_{t}$ jet algorithm~\cite{Cacciari:2008gp,Cacciari:2005hq,Cacciari:2011ma} with a radius parameter of 0.4, and these are referred to as ``small-$R$ jets''. Each topological cluster is calibrated to the electromagnetic energy scale prior to jet reconstruction~\cite{Aad:2009wy}. The reconstructed jets are then calibrated to the particle level by the application of a jet energy scale derived from simulation~\cite{newJES}. After energy calibration, jets are required to satisfy the $\pt >25\gev$ and $|\eta|<2.5$ selection. Quality criteria are imposed to identify jets arising from non-collision sources or detector noise and any event containing such a jet is removed~\cite{Run2-jet-cleaning}. Finally, to reduce the effect of pileup, an additional requirement is made on the jet vertex tagger (JVT) discriminant~\cite{aad:2015ina} for jets with $\pt < 60\gev$ and $ |\eta| < 2.4$.

Jets are tagged as containing a $b$-hadron via a multivariate $b$-tagging algorithm~\cite{Aad:2015ydr,ATL-PHYS-PUB-2016-012}. For each jet, a value for the multivariate $b$-tagging discriminant is calculated, and the jet is considered $b$-tagged if this value is above a given threshold. The threshold used in this search corresponds to an average 77\% efficiency to tag a jet containing a $b$-hadron, with a light-jet rejection factor of $\sim$134 and a charm-jet rejection factor of $\sim$6, as determined for jets with $\pt >20\gev$ and $|\eta|<2.5$ in simulated $\ttbar$ events~\cite{Aaboud:2018xwy,ATLAS-CONF-2018-001,ATLAS-CONF-2018-006}.

To avoid assigning a single detector response to more than one reconstructed object, a sequential overlap-removal procedure is adopted. Electron candidates that lie within $\Delta R = 0.01$ of a muon candidate are removed to suppress contributions from bremsstrahlung. To prevent double-counting of electron energy deposits as jets, the closest jet within ${\Delta}R_{y} = \sqrt{ (\Delta y)^2 + (\Delta \phi)^2 } = 0.2$ of a selected electron is removed.\footnote{The rapidity is defined as $ y = \frac{1}{2} \ln \frac{E + p_{z}}{E - p_{z}} $, where $E$ is the energy and $p_{z}$ is the longitudinal component of the momentum along the beam pipe.} If the nearest jet surviving that selection is within ${\Delta}R_{y} = 0.4$ of an electron, the electron is discarded. The overlap removal procedure between the remaining jet candidates and muon candidates is designed to remove those muons that are likely to have arisen in the decay chain of hadrons and to retain the overlapping jet instead. Jets and muons may also appear in close proximity. For example, a muon with high-$\pt$ bremsstrahlung radiation inside the calorimeter may be reconstructed as a jet. In such cases, the jet should be removed and the muon retained. Such jets are characterized by having very few matching inner-detector tracks. Selected muons that satisfy $\Delta R(\mu,\textrm{jet}) < 0.04+10\gev/\pt^\mu$ are rejected if the jet has at least three tracks originating from the primary vertex; otherwise the jet is removed and the muon is kept.

The selected and calibrated small-$R$ jets with $\pt>25\gev$ and passing both the requirement on the JVT and the overlap removal with leptons are used as inputs for further jet reclustering~\cite{Nachman:2014kla} using the anti-$k_{t}$ algorithm with a radius parameter of $R=1.0$. These reclustered large-$R$ jets are referred to as RCLR jets. The calibration corrections and uncertainties in the \largeR jets are automatically inherited from the small-$R$ jets~\cite{ATLAS-CONF-2017-062}. In order to further suppress contributions from pileup and other soft radiation, the \largeR jets are trimmed~\cite{Krohn:2009th} by removing all small-$R$ jets within a reclustered jet that have $\pt$ below 5\% of the $\pt$ of the reclustered jet. Due to the pileup suppression and $\pt>25\gev$ requirements made on the small-$R$ jets,  the probability for a small-$R$ jet to be removed from the corresponding reclustered jet by the trimming requirement is less than 1\%. The resulting \largeR jets are used to identify hadronically decaying top-quark candidates. \largeR jets that have  $\pt > 200\gev$, $|\eta|<$~2.0, mass\footnote{The reclustered jet mass is computed from the sum of the four-momenta of the associated small-$R$ jets~\cite{ATLAS-CONF-2017-062}.} larger than $100\gev$ and at least one constituent small-$R$ jet are referred to as ``\mreclust\ jets''.

The missing transverse momentum in the event, whose magnitude will be denoted in the following by $\met$, is defined as the negative vector sum of the $\pt$ of reconstructed and calibrated objects in the event, where only primary objects enter the sum (e.g., \largeR jets are not used). This sum includes a term to account for energy from low-momentum particles in the event that are not associated with any of the selected objects, which  is calculated from inner detector tracks matched to the reconstructed primary vertex in the event~\cite{Aaboud:2018tkc}.

Events of interest are required to have at least one reconstructed lepton that matches, within $\Delta R < 0.15$, the lepton with the same flavor reconstructed by the trigger algorithm. Events in the opposite-sign dilepton channel are retained if they contain exactly two opposite-sign charged leptons (electrons or muons) and at least four jets satisfying the quality and kinematic criteria discussed above, of which at least two must be $b$-tagged. In both the $ee$ and $\mu\mu$ channels, the dilepton invariant mass ($m_{\ell\ell}$) must be above $50 \gev$ and outside the $Z$ mass window 83--99\gev. Events not in the opposite-sign dilepton channel may enter the single-lepton channel if they contain exactly one lepton and at least five jets, of which at least two are $b$-tagged. The above selection criteria imply that events containing two leptons with the same charge, or three or more leptons of any charge are excluded from the selection. This is done in order to maintain orthogonality with the complementary search with same-sign dilepton and multilepton final states~\cite{SS_4tops} carried out by ATLAS, as these results are combined with the results presented here (Section~\ref{sec:Results}). Additional requirements are made to suppress the background from multijet production in the single-lepton channel. Requirements are made on $\met$ as well as on the transverse mass of the combined lepton and $\met$ system\footnote{$\mtw = \sqrt{2 p^\ell_{\mathrm T} \met (1-\cos\Delta\phi)}$, where $p^\ell_{\mathrm T}$  is the transverse momentum of the lepton and $\Delta\phi$ is the azimuthal angle separation between the lepton and the direction of the missing transverse momentum.} ($\mtw$): $\met >20\gev$ and $\met +\mtw>60\gev$. 

The above requirements are referred to as ``preselection'' and are summarized in Table~\ref{tab:event_selection}. Events satisfying either the single-electron or single-muon selections are combined and treated as a ``single-lepton'' analysis channel, and events satisfying any of the opposite-sign lepton selections ($ee$, $\mu\mu$ or $e\mu$) are combined and treated as a ``dilepton'' analysis channel.

{
\begin{table*}[t!]
\centering
\caption{Summary of preselection requirements for the single-lepton and dilepton channels. Here $\mtw$ is the transverse mass of the lepton and the $\met$ vector, and $m_{\ell\ell}$ denotes the dilepton invariant mass in the $ee$ and $\mu\mu$ channels.}
\begin{tabular}{l c c}
\toprule\toprule
\multicolumn{3}{c}{Preselection requirements}  \\
\midrule
Requirement & Single-lepton &  Dilepton   \\
\midrule
Trigger & \multicolumn{2}{c}{Single-lepton triggers} \\
Leptons & 1 isolated   & 2 isolated, opposite-sign \\
Jets    & $\geq$5 jets   & $\geq$4 jets \\
$b$-tagged jets &   \multicolumn{2}{c}{$\geq$2 $b$-tagged jets}  \\
Other   & $\met >20\gev$       &  $m_{\ell\ell} > 50\gev$   \\
        & $\met +\mtw>60\gev$  &  $| m_{\ell\ell}  - 91\gev | > 8\gev$ \\
\bottomrule\bottomrule
\end{tabular}
\label{tab:event_selection}
\end{table*}
}

\section{Signal and background modeling}
\label{sec:SandB}
After the event preselection, the main background processes arise from the SM production of $\ttbar$+jets and single top-quarks, as well as $W$- or $Z$-boson production in association with jets. Small contributions arise from the associated production of a vector boson $V$ ($V = W,Z$) or a Higgs boson and a \ttbar\ pair (\ttV\ and \ttH) and from diboson ($WW$, $WZ$, $ZZ$) production. Multijet events contribute to the selected sample via the misidentification of hadronic objects (jets, hadrons) as leptons or the presence of a non-prompt electron or muon. These events are referred to as the ``fake and non-prompt lepton'' background in the remainder of this paper.

MC simulation samples are used to model the expected distributions of the signal and most of the background processes. The fake and non-prompt lepton background in the single-lepton channel is estimated with a fully data-driven method. The $\ttbar$+jets background, which is dominant in regions with very high jet and $b$-jet multiplicities, is estimated via a dedicated data-driven method, with some correction factors taken from the MC simulation, as described in Section~\ref{sec:ttDD}. The MC samples were processed either through the full ATLAS detector simulation \cite{Aad:2010ah} based on {\scshape Geant4}~\cite{Agostinelli:2002hh}, or through a faster simulation making use of parameterized showers in the calorimeters~\cite{ATL-PHYS-PUB-2010-013}. To model the effects of pileup, events from minimum-bias interactions were generated using the {\PYTHIA}~8.186~\cite{Sjostrand:2007gs} event generator and overlaid on the simulated hard-scatter events according to the luminosity profile of the recorded data. All simulated samples were processed through the same reconstruction algorithms and analysis chain as the data. In the simulation, the top-quark mass was assumed to be $m_{\textrm{top}} = 172.5\gev$. The heavy-flavor decays were modeled using the {\scshape EvtGen}~1.2.0~\cite{Lange:2001uf} program, except for processes modeled using the {\SHERPA} generator~\cite{gleisberg:2008ta}.

\subsection{Signal modeling}
\label{sec:Signal}
Simulated events for the main signal process, i.e.~the four-top-quark production with SM kinematics, were generated at leading order (LO) with the {\amcatnlo}~2.2.2~\cite{Alwall:2014hca} generator and the {\textsc NNPDF2.3} LO PDF set~\cite{Ball:2012cx}, interfaced to {\PYTHIA}~8.186 using the {\textsc A14} set of tuned parameters~\cite{ATLASUETune4}, which will be denoted in the following by {\textsc A14} tune. The SM $\tttt$ sample is normalized to a cross section of 9.2~fb, computed at NLO in QCD~\cite{Alwall:2014hca}.

This search also probes a BSM model with kinematic characteristics similar to those of the SM $\tttt$ events: the $\tttt$ production via an effective field theory involving a four-fermion contact interaction~\cite{degrande:2010kt}. The EFT $\tttt$ sample was generated at LO with the {\amcatnlo}~2.2.2 generator and the {\textsc NNPDF2.3} LO PDF set, interfaced to {\PYTHIA}~8.186 with the {\textsc A14} tune. It is normalized assuming $|C_{4t}|/\Lambda^2 = 4\pi$$\tev^{-2}$, where $C_{4t}$ denotes the coupling constant and $\Lambda$ the energy scale of new physics, which yields a cross section of 926.3~fb computed using {\amcatnlo}. Details of this BSM scenario can be found in Refs.~\cite{HTX,SS_4tops}.

\subsection{Background modeling}
\label{sec:bkgmodel_ttbar}
The dominant $\ttbar$+jets background estimation relies on the data-driven technique described in Section~\ref{sec:ttDD}. The validation of this technique and the extraction of the corresponding correction factors were performed with simulated MC $\ttbar$+jets events, generated with {\POWHEGBOX} v2~\cite{Frixione:2007nw,Nason:2004rx,Frixione:2007vw,alioli:2010xd}, which provides NLO accuracy in QCD for the \ttbar\ process and uses the {\textsc CT10} PDF set~\cite{lai:2010vv}. Showering was performed using {\PYTHIA}~6.428~\cite{sjostrand:2006za} with the CTEQ6L PDF set~\cite{pumplin:2002vw} and the {\textsc PERUGIA2012} tune~\cite{skands:2010ak}. The hard-process factorization scale $\mu_{\text{F}}$ and renormalization scale $\mu_{\text{R}}$ were set to the default {\Powheg} value: $\mu = \sqrt{m^{2}_{\textrm{top}} + p^{2}_{\textrm{T,top}}} $, where $p_{\textrm{T,top}}$ is the transverse momentum of the top quark in the $\ttbar$ center-of-mass reference frame. The {\Powheg} model resummation damping parameter, $h_\textrm{damp}$, which controls the matching of matrix elements to parton showers and regulates the high-$\pt$ parton radiation, was set to $m_{\textrm{top}}$~\cite{ATL-PHYS-PUB-2015-002}. The sample is normalized to the theoretical cross-section value for the inclusive \ttbar\ process of $832^{+40}_{-46}$~pb obtained with {\scshape Top++}~\cite{czakon:2011xx}, calculated at next-to-next-to-leading order (NNLO) in QCD, and including resummation of next-to-next-to-leading logarithmic soft gluon terms~\cite{xs1,xs2,xs3,xs4,xs5}.

Samples of $W/Z$+jets events were generated with the {\SHERPA}~2.2~\cite{gleisberg:2008ta} generator. The matrix element calculation was performed with up to two partons at NLO in QCD and up to four partons at LO using matrix elements from \textsc{Comix}~\cite{gleisberg:2008fv} and \textsc{OpenLoops}~\cite{cascioli:2011va}. The matrix element calculation was merged with the {\SHERPA}~\cite{schumann:2007mg} parton shower (PS) using the ME+PS@NLO prescription~\cite{hoeche:2012yf}. The PDF set used for the matrix element calculation is {\textsc NNPDF3.0nnlo} with a dedicated PS tuning developed by the {\SHERPA} authors. The $W$+jets and $Z$+jets samples are normalized to their inclusive production cross section estimates at NNLO in QCD, calculated with \textsc{FEWZ}~\cite{Melnikov:2006kv,Gavin:2010az}.

Samples of single-top-quark backgrounds, corresponding to the $Wt$ and $s$-channel production mechanisms, were generated with {\POWHEGBOX} v1~\cite{Alioli:2009je} at NLO accuracy using the {\textsc CT10} PDF set. Overlaps between the \ttbar\ and $Wt$ final states were removed using the ``diagram removal'' scheme~\cite{frixione:2005vw}. Samples of $t$-channel single-top-quark events were generated using the {\POWHEGBOX} v1~\cite{Re2010,frederix:2012dh} NLO generator that uses the four-flavor scheme. The fixed four-flavor PDF set \textsc{CT10}f\textsc{4}~\cite{lai:2010vv} was used for the matrix element calculations. Showering was performed using {\PYTHIA}~6.428 with the {\textsc PERUGIA2012} tune. The single-top-quark samples are normalized to the approximate NNLO cross sections~\cite{kidonakis:2011wy,kidonakis:2010ux,kidonakis:2010tc}.

Diboson processes with one of the bosons decaying hadronically and the other leptonically were simulated using the {\SHERPA}~2.1.1 generator. They were calculated for up to one ($ZZ$) or zero ($WW$, $WZ$) additional partons at NLO, and up to three additional partons at LO, using the same procedure as for $W/Z$+jets. The {\textsc CT10} PDF set was used together with a dedicated PS tuning of the {\Sherpa} fragmentation model. All diboson samples are normalized to their NLO cross sections provided by {\SHERPA}.

Samples of $\ttbar+V$ (with $V$ = $W$ or $Z$, including non-resonant $Z/\gamma^{*}$ contributions) were generated with {\amcatnlo}~2.3.2, using NLO in QCD matrix elements and the {\textsc NNPDF3.0NLO}~\cite{Ball:2014uwa} PDF set. Showering was performed using {\pythia}~8.210 and the {\textsc A14} tune. The $\ttbar+V$ events are normalized to their NLO cross section~\cite{Alwall:2014hca}. A sample of $\ttbar+H$ events was generated using {\amcatnlo}~2.3.2 generator and the {\textsc NNPDF3.0NLO} PDF set. Showering was performed using {\PYTHIA}~8.210 and the {\textsc A14} tune. Inclusive decays of the Higgs boson are assumed in the generation of the $\ttbar+H$ sample, which is normalized to the corresponding cross section calculated at NLO~\cite{dittmaier:2011ti,dittmaier:2012vm}. Rare backgrounds, such as $\ttbar+WW$ and triple-top-quark production ($\ttbar+t$, $\ttbar+tW$), were generated at LO with {\amcatnlo}~2.2.2 with no additional partons and interfaced with {\pythia}~8.186. They are normalized using cross sections computed at NLO in QCD~\cite{Alwall:2014hca,Barger:2010uw}.

\subsection{Estimation of non-prompt and fake lepton backgrounds}
\label{sec:fake_lj}
In the single-lepton channel, the background from events with a fake or non-prompt lepton is estimated from data using a ``matrix method'' technique ~\cite{aad:2010ey,atlas-conf-2014-058}. Events are selected using looser isolation or identification requirements for the lepton and are then weighted according to the efficiencies for both prompt and background (fake and non-prompt) leptons to pass the tighter default selection. These efficiencies are measured in data using dedicated control regions. The contribution from events with a fake or non-prompt lepton is found to be consistent with zero in regions defined by the presence of two or more \mreclust\ jets, as well as in the regions requiring the presence of at least one \mreclust\ jet and at least four $b$-tagged jets. The contribution is at most 6\% in the rest of the signal regions (described in Section~\ref{sec:AnalysisMethod}). 

In the dilepton channel, the majority (90\%) of events containing one prompt lepton and one background lepton, arising from either a heavy-flavor hadron decay, photon conversion, jet misidentification or light-meson decay, originate from the single-lepton $\ttbar$+jets background. This contribution is included in the estimation via the data-driven technique described in Section~\ref{sec:ttDD}, while the small fraction of fake and non-prompt leptons events arising from $W$+jets and $\ttbar+V$ events is estimated from MC simulation.\footnote{No data-driven estimation was attempted due to the statistical overlap with the same-sign dilepton and multilepton final-states search~\cite{SS_4tops}.} The total contribution is found to be less than 8\% of the total background in the signal regions.

\section{Search strategy}
\label{sec:AnalysisMethod}
\label{sec:Classify}
Signal events from SM four-top-quark production in the single lepton (opposite-sign dilepton) decay channel are characterized by the presence of one charged lepton (two opposite-sign charged leptons), missing transverse momentum from the escaping neutrino(s) and a high number of  high-$\pt$ jets. At LO the single-lepton (opposite-sign dilepton) decay will potentially have an event topology with ten (eight) jets, when each parton from a top-quark decay gives rise to a separate jet: six (four) jets are light-jets and four are $b$-quark jets. However, the topology of a reconstructed event could differ due to the limited detector acceptance, the $b$-tagging efficiency, and the possible presence of jets arising either from additional radiation and multiple parton interactions (MPI) or from collimated partons not resolved as separate objects. Events are classified in several regions to optimize the sensitivity of the search, to perform a data-driven estimate of the $\ttbar$+jets background (described in Section~\ref{sec:ttDD}) and to validate the background prediction.

Preselected events in each of the two channels are classified according to their event topology, defined by the number of jets with $\pt > 25\gev$ and the number of $b$-jets. Several regions are split according to the \mreclust\ jet multiplicity in addition to the jet and $b$-tagged jet multiplicities. In the following, a region with $m$ jets (j), of which $n$ are $b$-tagged jets (b) and from which $p$ separate \mreclust\ jets (J) are reconstructed is referred to as ``$m$j, $n$b, $p$J''. When no \mreclust\ jet multiplicity is specified, no selection on these objects is performed.

\begin{figure}[h]
\centering
\subfigure[]{\includegraphics[height=0.22\textheight]{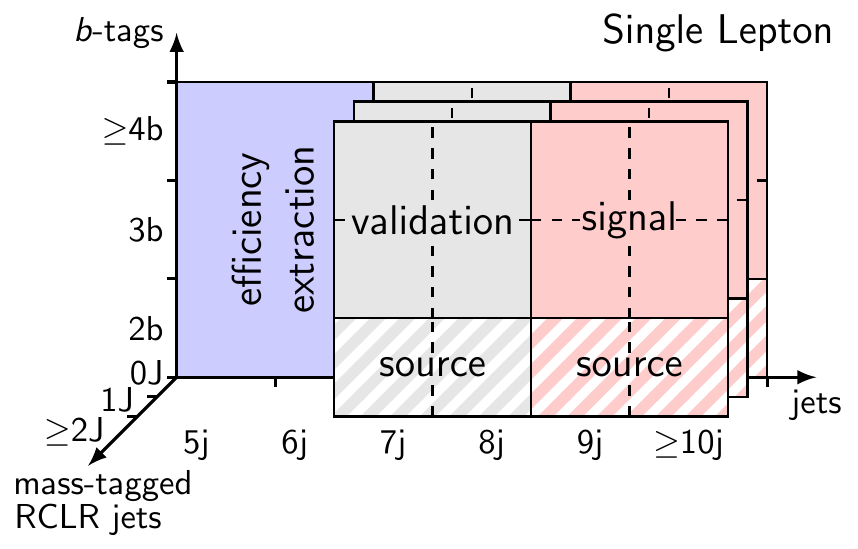}}
\subfigure[]{\includegraphics[height=0.22\textheight]{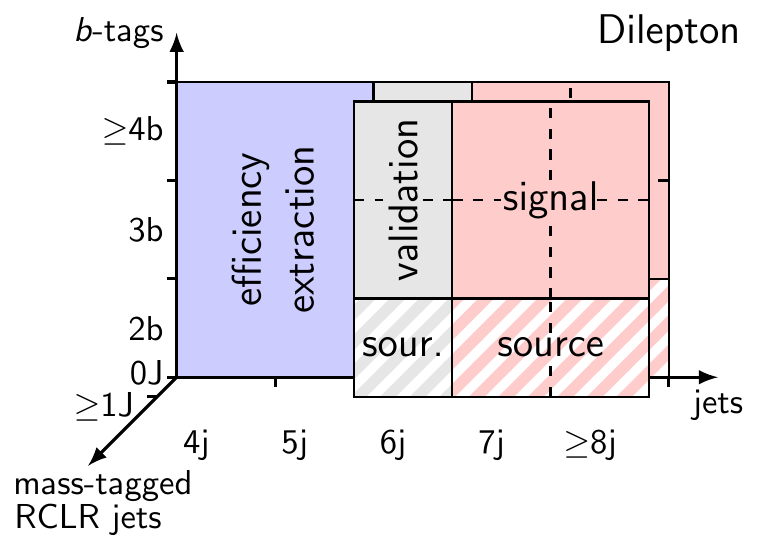}}
\caption{
Schematic view of the different analysis regions in (a) the single-lepton and (b) the dilepton channels. The three axes represent the jet multiplicity, the $b$-tagged jet multiplicity and the \mreclust\ jet multiplicity. The efficiency extraction region in each channel is defined inclusively in the \mreclust\ jet multiplicity.
}
\label{fig:ttDD_regions}
\end{figure}

The following regions are defined to be orthogonal using the classification described above: 20 ``signal regions,'' 16 ``validation regions,'' 18 ``source regions'' and 2 ``efficiency extraction regions,'' as shown in Figure~\ref{fig:ttDD_regions}.

Twelve regions in the single-lepton channel and eight regions in the dilepton channel with the largest signal-to-background ratios (up to 5.7\% in the single-lepton channel and 7.0\% in the dilepton channel), assuming SM \tttt\ production cross section and kinematics, are referred to as signal regions. These regions are included in the simultaneous fit to extract the signal cross section and have high jet multiplicities ($\geq$9j and $\geq$7j for single-lepton and dilepton respectively) and high $b$-tagged jet multiplicities ($\geq$3b). Since events from the main $\ttbar$+jets background are characterized by at most one hadronically decaying top quark in the single-lepton channel and no hadronically decaying top quarks in the dilepton channel, the signal regions are split into 0, 1 and $\geq$2J in the single-lepton case, and into 0 and $\geq$1J in the dilepton case.

Twelve validation regions in the single-lepton channel and four validation regions in the dilepton channel are defined. These regions do not overlap with the signal region selections and feature low expected signal-to-background ratios (less than 1\%). They are not included in the fit nor used to extract information from the data. These regions are designed primarily to validate the data-driven estimate of the $\ttbar$+jets background (introduced in Section~\ref{sec:ttDD}) and to confirm the validity of the assumption that the $\ttbar$+jets data-driven estimate can be extrapolated to the signal regions. The validation regions in the single-lepton channel contain exactly seven or exactly eight jets of which three or at least four are $b$-tagged. In the dilepton channel, the validation regions have exactly six jets of which three or at least four are $b$-tagged. In each of the two channels these validation regions are split according to the \mreclust\ jet multiplicity in the same way as the corresponding signal regions.

With the goal of estimating the $\ttbar$+jets background in the signal regions, data events with lower jet and/or $b$-jet multiplicities are used in the data-driven method described in Section~\ref{sec:ttDD}. The 18 source regions are built using events with high jet multiplicity: 7, 8, 9, $\geq$10 for the single-lepton channel and 6, 7, $\geq$8 for the dilepton channel, out of which exactly 2 jets are $b$-tagged. They are used to build pseudo-data event samples in the signal and validation regions with same jet multiplicities but higher number of $b$-tagged jets. Efficiency extraction regions are characterized by lower jet multiplicities: five or six jets for the single-lepton channel and four or five for the dilepton channel, out of which 2, 3 or $\geq$4 are $b$-tagged. They are used to extract the $b$-tagging probabilities, since they provide a sample depleted of signal and dominated by $\ttbar$+jets. Neither the efficiency extraction regions nor the source regions are included in the final fit to data.

\begin{figure}[ht]
\centering
\subfigure[]{\includegraphics[width=0.4\textwidth]{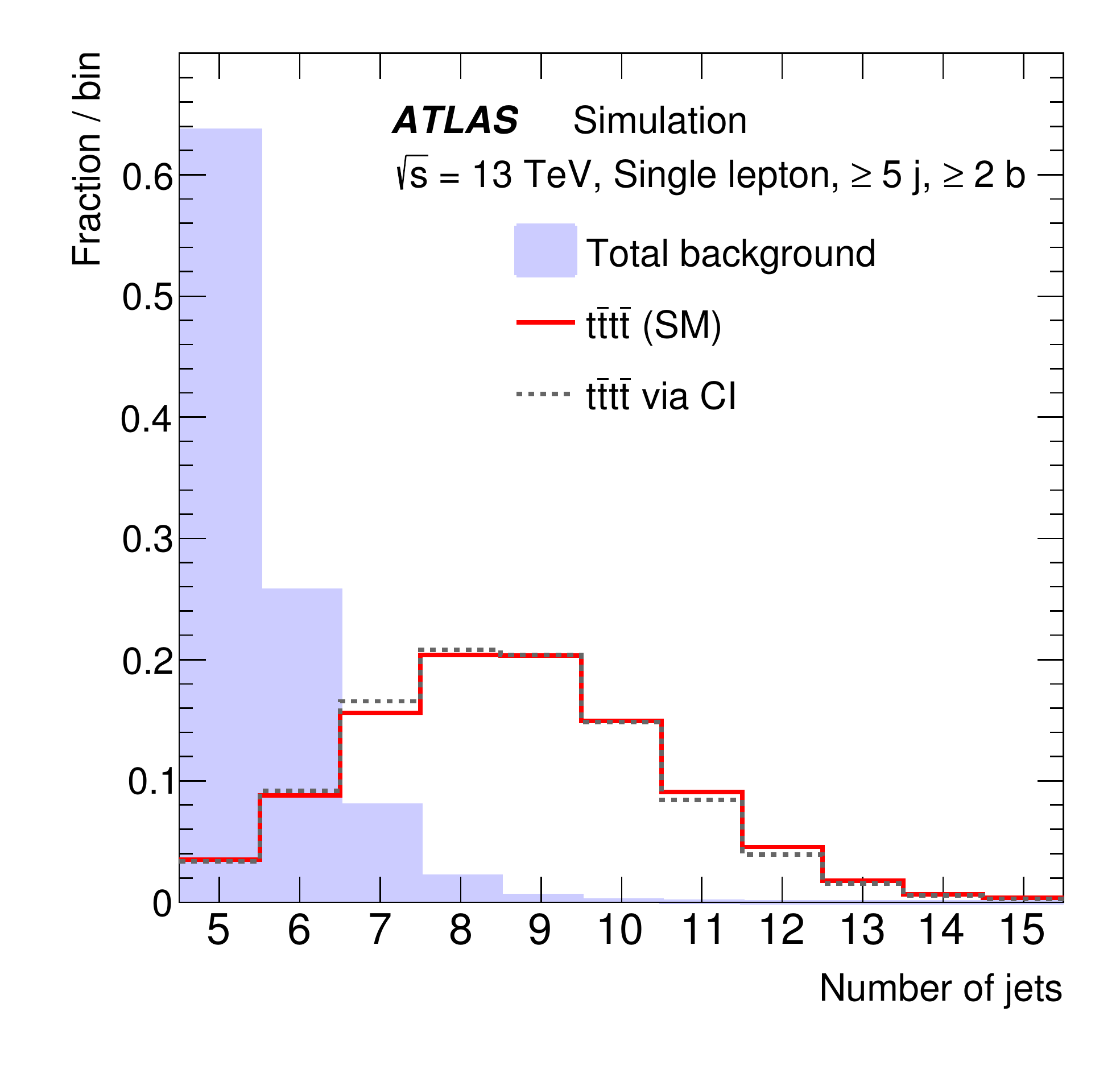}}  
\subfigure[]{\includegraphics[width=0.4\textwidth]{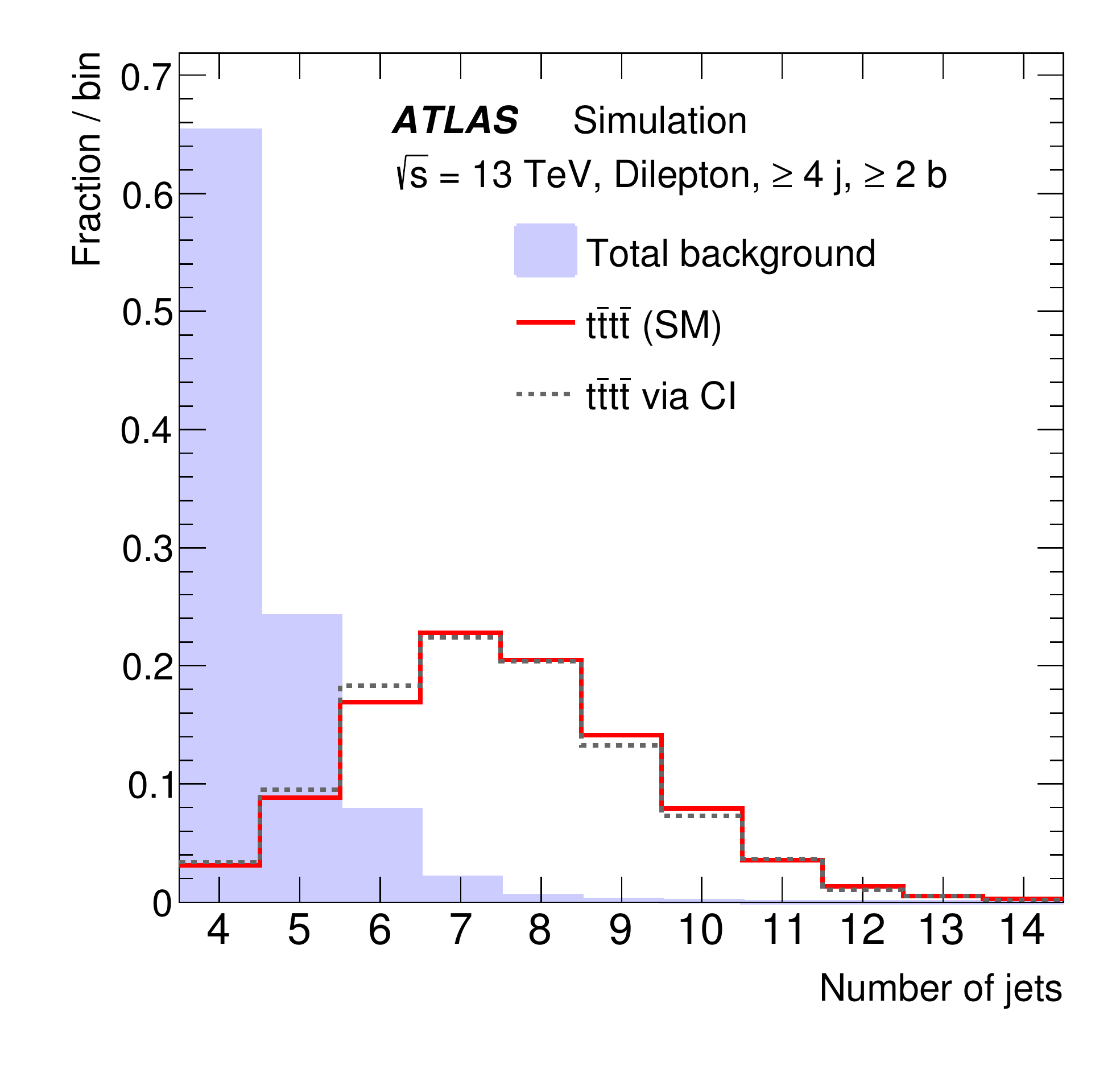}} \\
\subfigure[]{\includegraphics[width=0.4\textwidth]{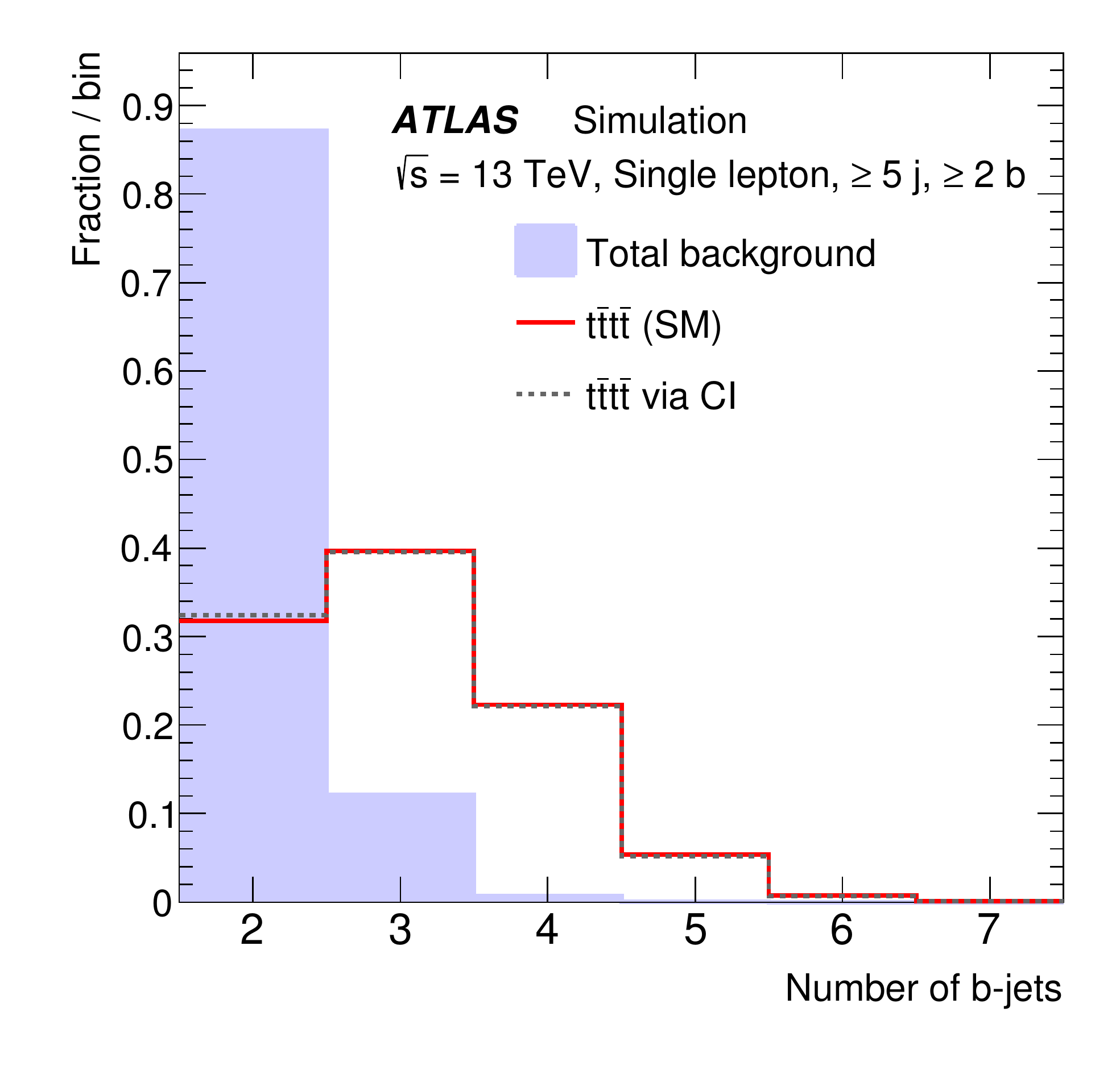}}  
\subfigure[]{\includegraphics[width=0.4\textwidth]{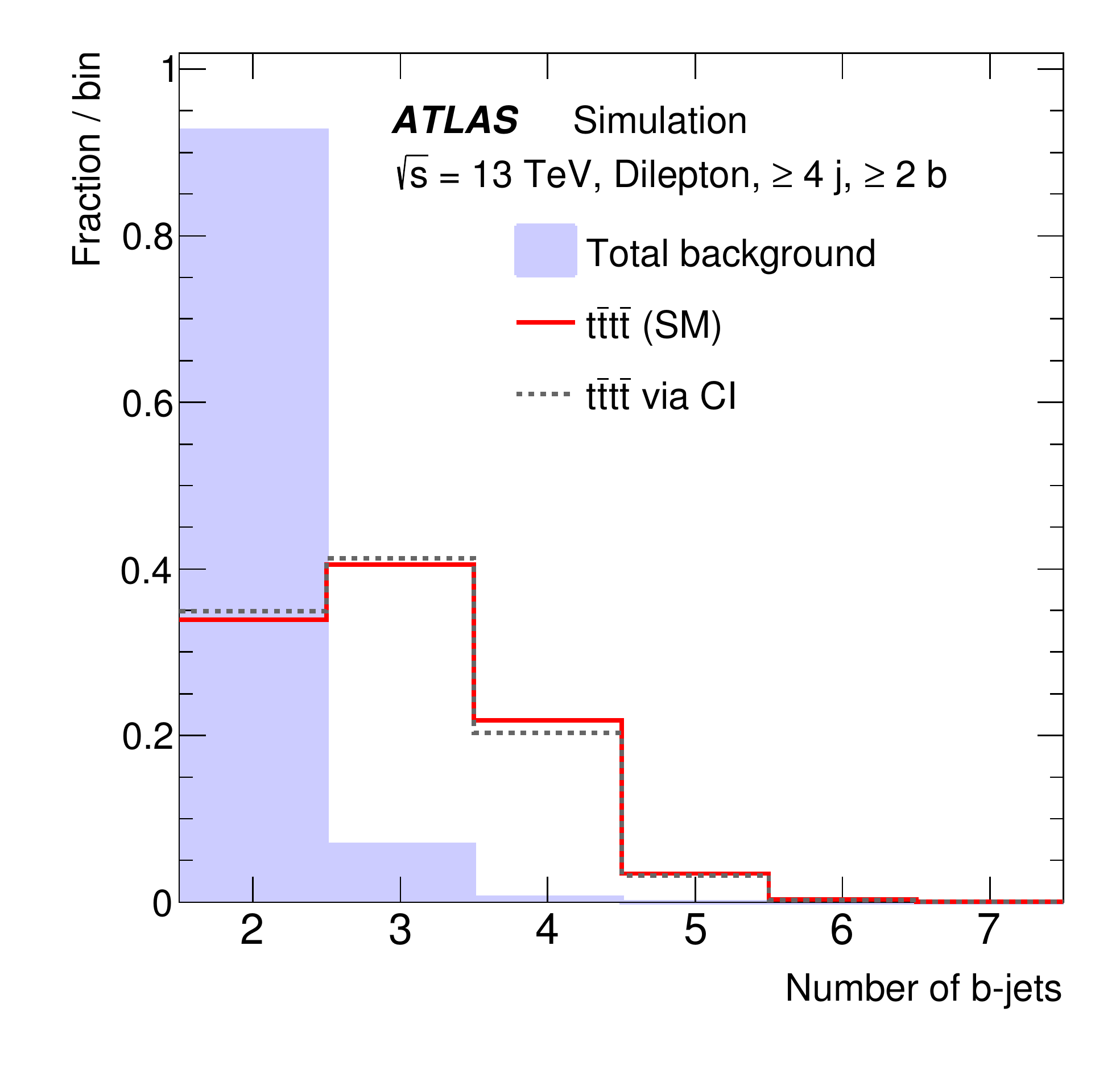}}
\caption{(a, b) The jet multiplicity and (c, d) the $b$-jet multiplicity distributions after preselection for the total predicted background with the $\ttbar$+jets background estimated via MC simulation (shaded histogram) and the signal scenarios considered in this search in the single-lepton (a, c) and the dilepton (b, d) channels. The signals shown correspond to four-top-quark production with SM kinematics (solid) and $\tttt$ production involving a four-fermion contact interaction (dashed). The distributions are normalized to unit area. The last bin in each distribution contains the overflow.
}
\label{fig:sep}
\end{figure}

\begin{figure}[ht]
\centering
\subfigure[]{\includegraphics[width=0.4\textwidth]{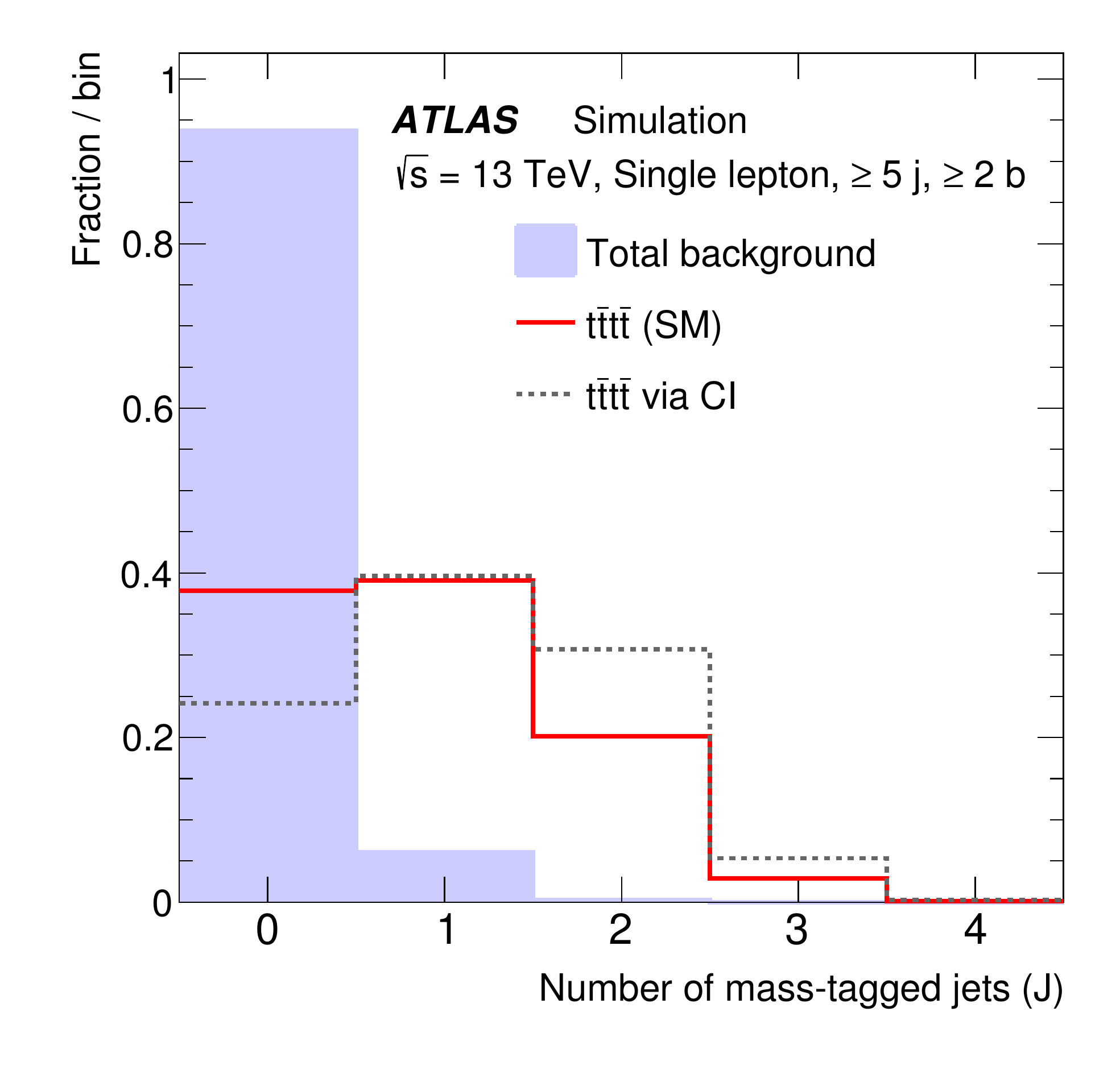}}  
\subfigure[]{\includegraphics[width=0.4\textwidth]{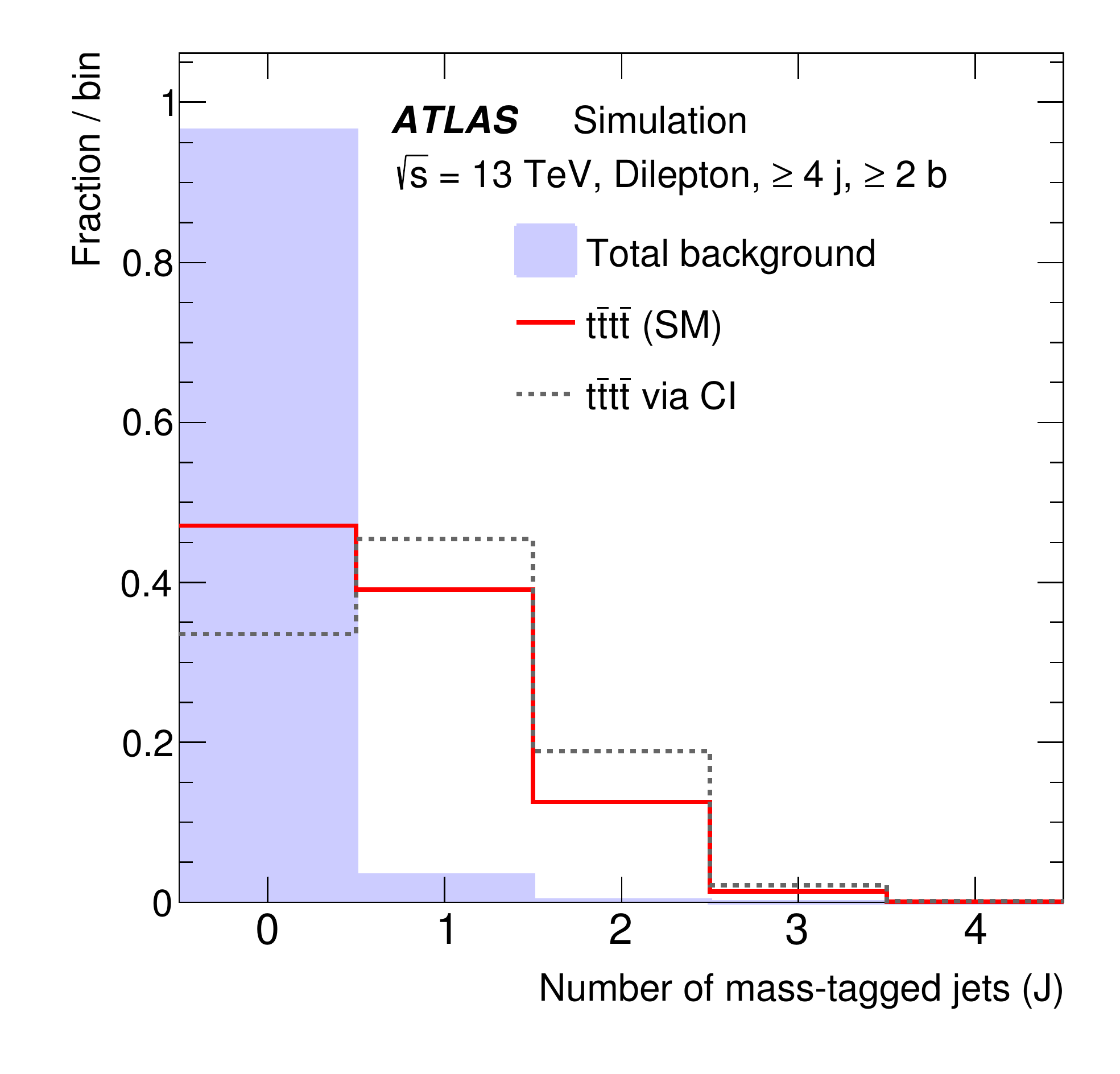}} \\
\subfigure[]{\includegraphics[width=0.4\textwidth]{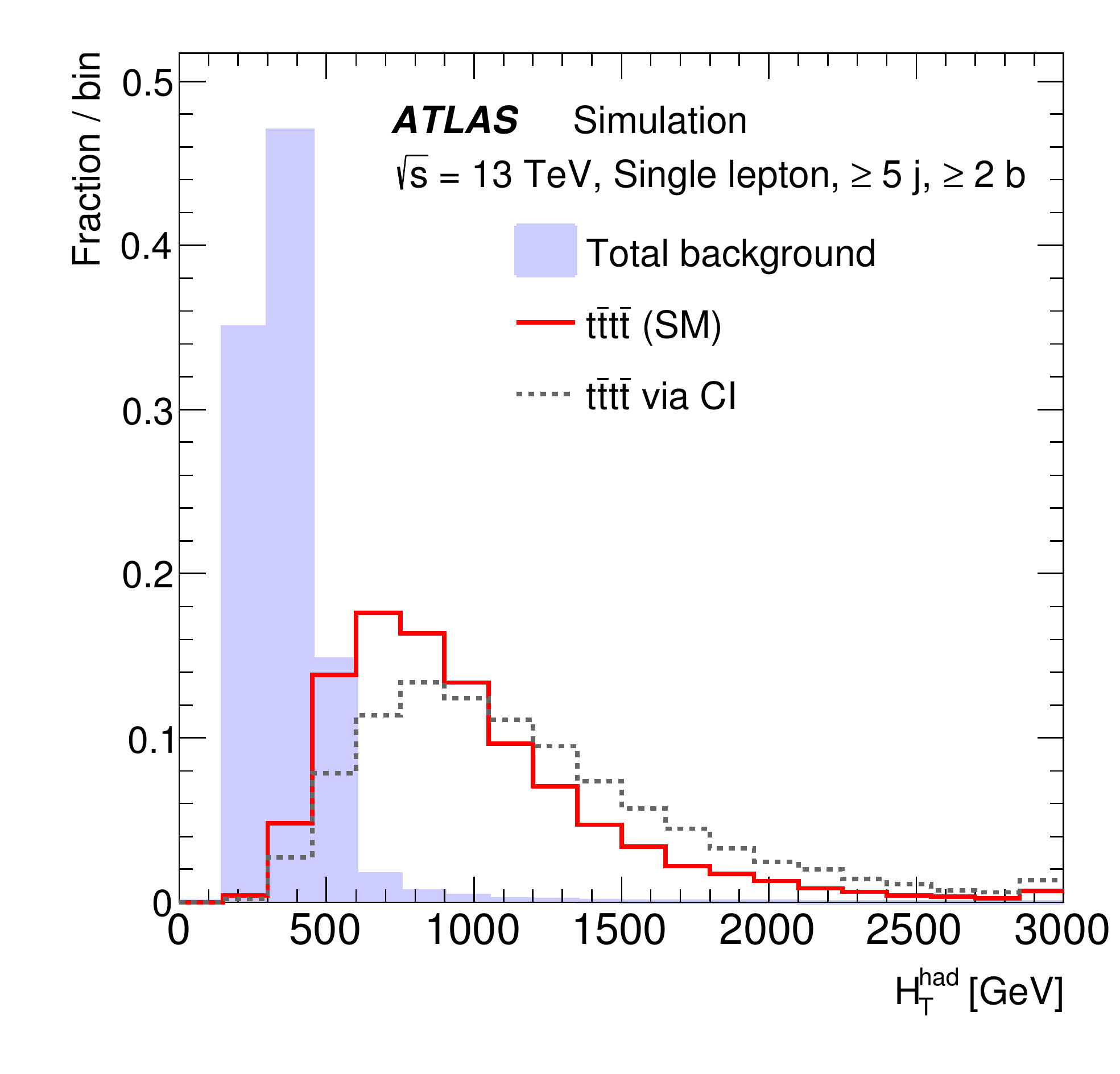}}  
\subfigure[]{\includegraphics[width=0.4\textwidth]{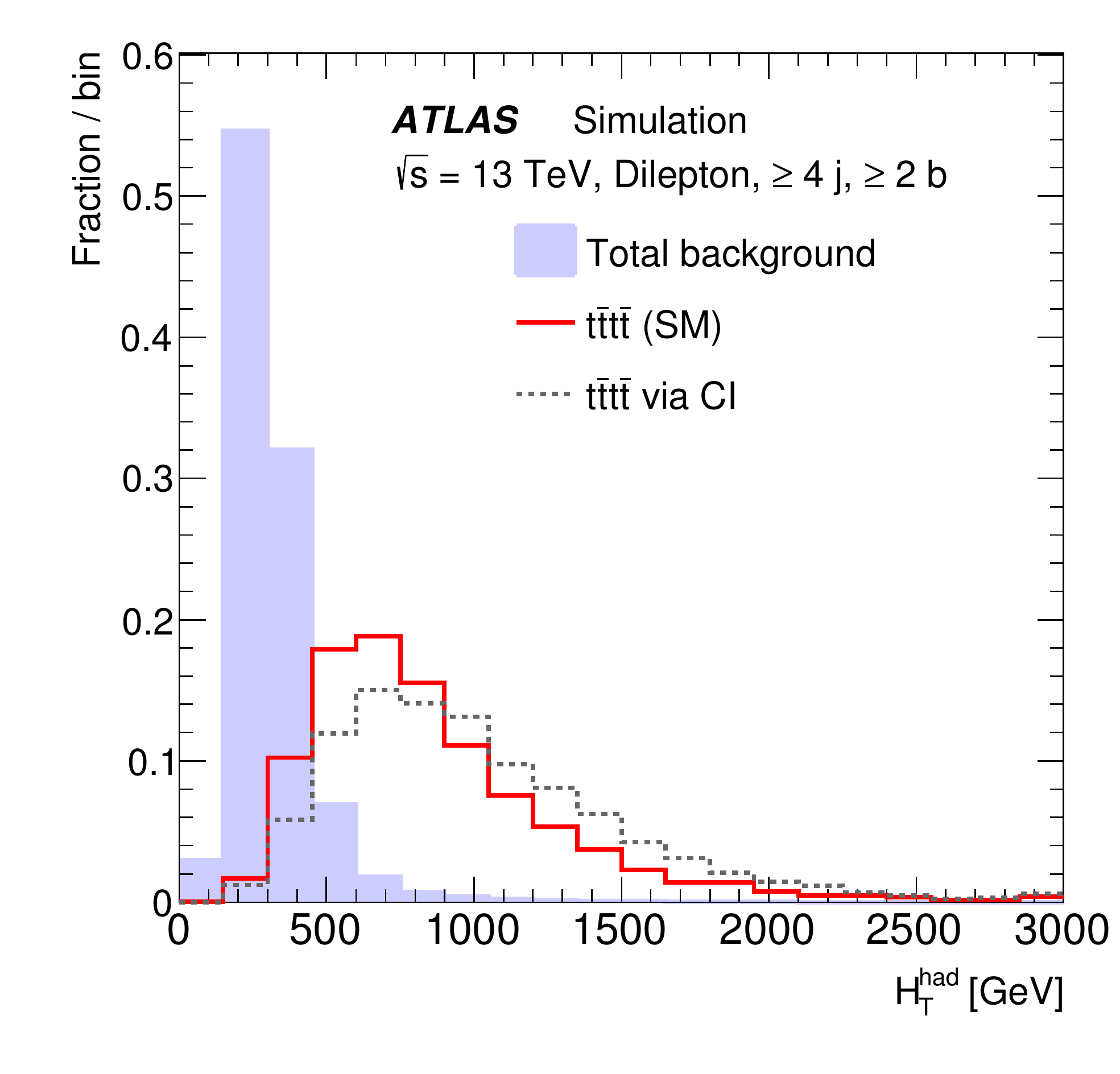}}
\caption{(a, b) The \mreclust\ jet multiplicity distributions and (c, d) the $\hthad$ distributions after preselection for the total predicted background with the $\ttbar$+jets background estimated via MC simulation (shaded histogram) and signals for the single-lepton (a, c) and the dilepton (b, d) channels. The signals shown correspond to four-top-quark production with SM kinematics (solid) and $\tttt$ production involving a four-fermion contact interaction (dashed). The distributions are normalized to unit area. The last bin in each distribution contains the overflow.
}
\label{fig:sep2}
\end{figure}

Figure~\ref{fig:sep} shows the expected shapes of the jet and $b$-jet multiplicity distributions after preselection in the single-lepton and dilepton channels. The distributions shown are for the total predicted background, with the $\ttbar$+jets background estimated via MC simulation, and for the considered four-top-quark signal scenarios. Figures~\ref{fig:sep2}(a) and \ref{fig:sep2}(b) show the same distributions but for the \mreclust\ jet multiplicity. 

Figures~\ref{fig:sep2}(c) and \ref{fig:sep2}(d) compare the expected shapes of the scalar sum of the jet transverse momenta, considering all selected jets ($\hthad$), between the different four-top-quark signal scenarios and the total predicted background. Given the different kinematic features, the $\hthad$ distribution provides a suitable discrimination between events from the signal hypotheses and the background, and is used as the main discriminating variable in each of the regions. The signal-to-background discrimination is therefore provided by the combination of the event categorization and the $\hthad$ distribution in each category.

\pagebreak
\clearpage

\section{$\ttbar$+jets background estimation using data: the TRF$_{\ttbar}$ method}
\label{sec:ttDD}
\label{sec:validationpre-fit} 
The MC simulation-based approach at NLO accuracy in QCD for the prediction of the inclusive $\ttbar$ background is not expected to model well the very high jet and $b$-jet multiplicity regions exploited in this search. Given the lack of multi-leg calculations, the MC simulation-based approach relies on the description of such large multiplicities through the parton-shower formalism with consequently large uncertainties. Therefore, a data-driven method is used to estimate the dominant background from $\ttbar$+jets in regions with very high jet and $b$-jet multiplicities. This method provides a more accurate prediction of this background than a purely simulation-based approach and avoids the need to estimate modeling uncertainties (documented in Section~\ref{sec:Systematics}) by extrapolation from kinematic regimes with different numbers of jets and $b$-tagged jets.

The estimate is based on a method introduced in Ref.~\cite{bertella:2059786} and is referred to as ``tag rate function for $\ttbar$+jets events'', which will be denoted in the following  by TRF$_{\ttbar}$. The method assumes that the probability of $b$-tagging an additional\footnote{Additional refers to all jets in addition to the $b$($\bar{b}$)-jets originating from the $\ttbar$ decay. This includes the jets possibly originating from hadronically decaying $W$ bosons.} jet in a $\ttbar$+jets event, where the additional jets can include $c$-~and~$b$-jets, is essentially independent of the number of additional jets. With this assumption, the tagging probability, as a function of the kinematic properties of the jet, can be estimated in lower jet-multiplicity events and then applied to data events with the same jet multiplicity as signal-region events, but lower $b$-tagged jet multiplicity, where the signal contamination is negligible. These $b$-tagging probabilities are measured and applied as a function of some of the jet and event properties. Simulation-based corrections are then applied in order to correct for the fact that the assumptions stated above may not be completely valid. 
Systematic uncertainties in these corrections are propagated through the final estimate.

The per-jet $b$-tagging probabilities $\varepsilon_b$ are measured in the efficiency extraction regions (described in Section~\ref{sec:AnalysisMethod}), after subtracting the contribution from all non-\ttbar\ processes modeled with MC simulation, amounting to 8--14\% of the total background, depending on the channel and on the signal region considered. In order to take into account the correlation of $\varepsilon_b$ with the $b$-tagged jet multiplicities, two sets of probabilities $\varepsilon_b^{\geq2b}$ and $\varepsilon_b^{\geq3b}$ are extracted separately for each of the two analysis channels. The measurement of $\varepsilon_b^{\geq2b}$ ($\varepsilon_b^{\geq3b}$) is done from events with $\geq$2 ($\geq$3) $b$-tagged jets. The two (three) $b$-tagged jets with the highest values of the multivariate $b$-tagging discriminant in the event are excluded from the computation. All probabilities $\varepsilon_b$ are measured both as a function of jet $\pt$ and as a function of the quantity $\Delta R^{\textrm{jet,jet}}_{\mathrm{min}} \times N^{\mathrm{jet}}$: the minimum distance in the $\eta$--$\phi$ plane between the given jet and all the other jets in the event, multiplied by the jet multiplicity\footnote{Assuming a uniform random distribution of jets across the $\eta$--$\phi$ plane, $\Delta R^{\textrm{jet,jet}}_{\mathrm{min}}$ is inversely proportional to $N^{\mathrm{jet}}$. Variables parameterizing the $b$-tagging probability should be chosen to be mostly independent of $N^{\mathrm{jet}}$, to allow the extrapolation of the $b$-tagging probabilities from low to high multiplicity regions.} $N^{\mathrm{jet}}$, chosen in order to take into account the correlation between the $b$-tagging probability and the presence of nearby jets (see Ref.~\cite{bertella:2059786}).

\begin{figure}[ht]
\centering
\includegraphics[width=0.75\textwidth]{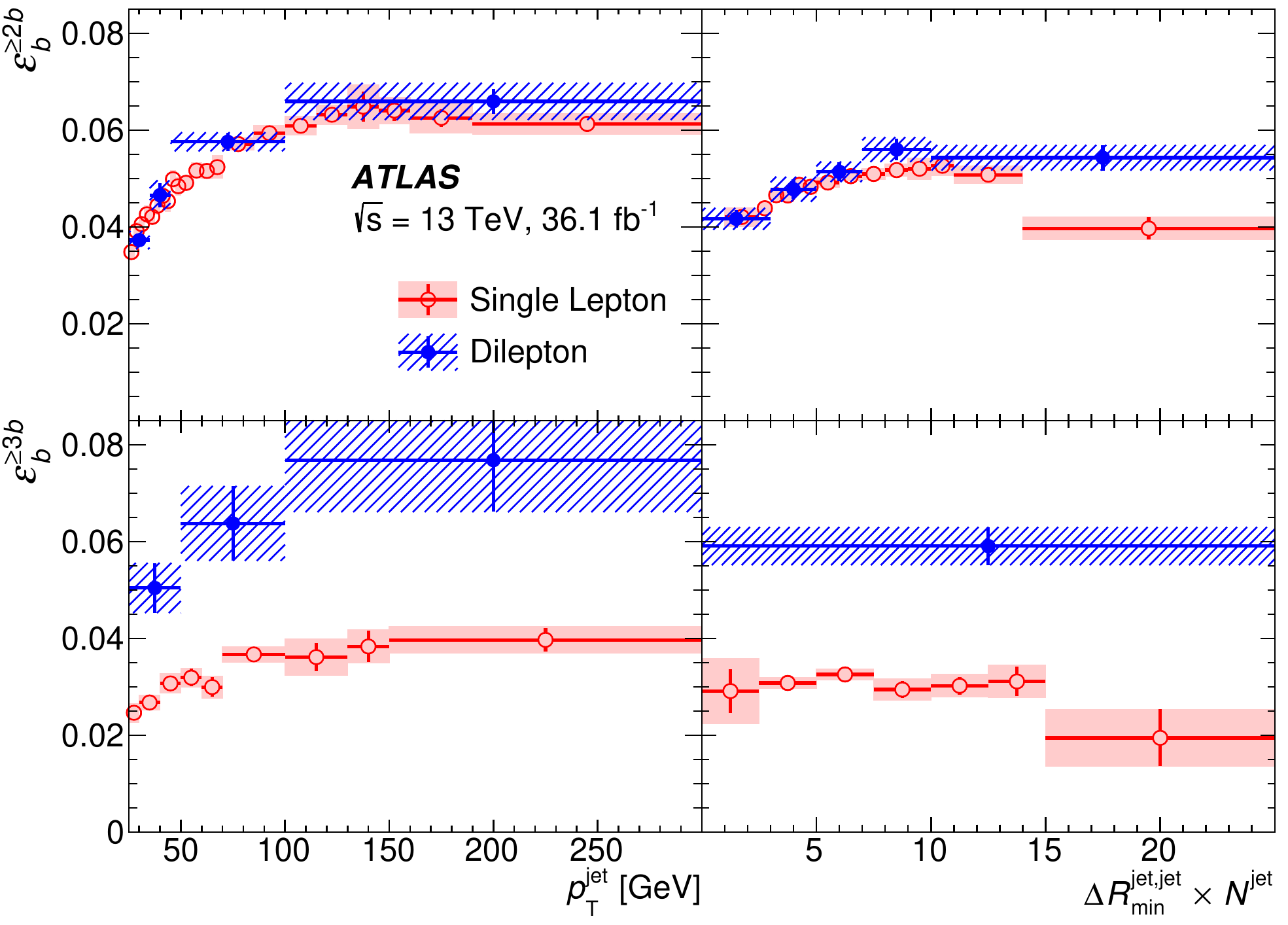}
\caption{Values of the per-jet $b$-tagging probability for $\ttbar$+jets events as a function of the jet transverse momentum ($\pt^{\textrm{jet}}$) and the minimum $\Delta R$ between the considered jet and the other jets in the event, scaled by the jet multiplicity ($\Delta R^{\textrm{jet,jet}}_{\mathrm{min}} \times N^{\textrm{jet}}$), as measured in data requiring the presence of at least two $b$-jets ($\varepsilon_b^{\geq2b}$) and least three $b$-jets ($\varepsilon_b^{\geq3b}$). $b$-tagging probabilities are shown separately for single-lepton and dilepton events. The vertical error bars indicate the statistical uncertainties, while the shaded (hatched) areas indicate the combined statistical and systematic uncertainties.  In the dilepton case, a constant $b$-tagging probability is assumed as a function of $\Delta R^{\textrm{jet,jet}}_{\mathrm{min}} \times N^{\textrm{jet}}$ for $\varepsilon_b^{\geq3b}$.
}
\label{fig:ttDD_eff}
\end{figure}

Figure~\ref{fig:ttDD_eff} shows the measured values of the $b$-tagging probability in the single-lepton and dilepton channels. It can be seen how the $\varepsilon_b^{\geq3b}$ are systematically lower than the $\varepsilon_b^{\geq2b}$ in the case of the single-lepton channel, while they are systematically higher in the case of the dilepton channel. This effect is due to the presence of hadronically decaying $W$ bosons only in the single-lepton channel, which can give rise only to light-jets or $c$-jets. In the dilepton case, when $\varepsilon_b^{\geq3b}$ is computed in the dominant four jet multiplicity,  this leaves only one jet where this $b$-tagging probability can be sampled, and this jet is likely to be a $b$-jet or $c$-jet, neglecting the mis-tag probability and considering the relative contributions of $\ttbar$+single and double $c/b$ through gluon splitting. This is not the case in the single-lepton channel, where, instead, three tagged jets out of five can easily be the consequence of tagging a $c$-jet from the $W$ boson, hence reducing the probability of tagging an additional jet. In the dilepton case, the dependence on $\Delta R^{\textrm{jet,jet}}_{\mathrm{min}} \times N^{\textrm{jet}}$ for the $\geq$3$b$ selection was found to be compatible with a constant value within statistics.

These $b$-tagging probabilities are then used to build ``pseudodata samples'' in validation and signal regions: this is done by applying the information derived from the measured $\varepsilon_b$ to the data in the source regions containing the same number of jets and \mreclust\ jets, accounting for the fact that this starting sample contains two $b$-tagged jets~\cite{bertella:2059786}. The small non-$\ttbar$+jets background contribution is subtracted, analogously to the procedure described in Ref.~\cite{abazov:2006ka}. In this way, jets that were not $b$-tagged in the original data sample can be promoted to $b$-tagged jets in a given pseudodata events sample, with a weight determined by $\varepsilon_b$, which accounts for the corresponding probability. For the estimate in the $3b$ categories, the procedure above is applied using only $b$-tagging probabilities extracted from events in the $\geq$2$b$ region ($\varepsilon_b^{\geq2b}$). For the estimate in the $\geq$4$b$ categories, a two-step procedure is applied: the estimates in the corresponding $3b$ categories are used as the starting point to apply again the same procedure, now using $b$-tagging probabilities extracted from events in the $\geq$3$b$ region ($\varepsilon_b^{\geq3b}$).

The last step of the method relies on the MC simulation to correct the estimate in each of the considered bins and to assign a set of systematic uncertainties. In order to achieve this, all the steps described above are applied to MC simulated $\ttbar$+jets events: the $b$-tagging probability $\varepsilon_b$ is extracted from simulated events in the efficiency extraction regions and is then used to reweight simulated events in the source regions, obtaining an estimate in the signal and validation regions. The resulting estimate in bin $i$ of \hthad\ distributions, $B^{\mathrm{TRF}_{\ttbar},\mathrm{MC}}_i$, is then compared with the prediction from simulated events selected in the signal and validation regions, $B^{\mathrm{MC}}_i$, extracting a correction factor for each considered bin, defined as $C_i = B^{\mathrm{MC}}_i/B^{\mathrm{TRF}_{\ttbar},\mathrm{MC}}_i$. The correction is then applied bin-by-bin to the purely data-driven estimate, $B^{\mathrm{TRF}_{\ttbar},\mathrm{Data}}_i$ to obtain a corrected estimate in each bin, $B^{\mathrm{TRF}_{\ttbar} }_i = C_i \times B^{\mathrm{TRF}_{\ttbar},\mathrm{Data}}_i$. These corrections reweight $B^{\mathrm{TRF}_{\ttbar},\mathrm{Data}}_i$ by less than $20\%$ on average, varying in magnitude region by region, and are primarily aimed to account for effects such as the dependence of the $b$-tagging probability on other jet or event properties than the ones used in the parametrization.

A full set of systematic uncertainties is then derived for the estimate $B^{\mathrm{TRF}_{\ttbar} }_i$ by repeating the described procedure on MC simulated events with systematic variations applied. For each considered source of systematic uncertainty affecting the $\ttbar$+jets MC prediction (see Section~\ref{sec:Systematics}), a new set of correction factors $C'_i$ is derived. In this ratio, systematic variations $\Delta B_i$ partially cancel out since $C’_i= (B+\Delta B)^{\mathrm{MC}}_i/(B+\Delta B)^{\mathrm{TRF}_{\ttbar},\mathrm{MC}}_i \simeq C_i \times [ 1+(\Delta B_i/B_i)^{\mathrm{MC}} - (\Delta B_i/B_i)^{\mathrm{TRF}_{\ttbar},\mathrm{MC}} ]$. The cancellation is exact for some uncertainties, e.g. overall normalization. Besides the systematic uncertainties, two sources of statistical uncertainties are considered. The first is the statistical uncertainty affecting the purely data-driven estimate, due to the limited numbers of data events in the source regions. The second source comes from the MC correction factor, given the limited number of simulated events both in the source regions and in the signal and validation regions.

\begin{figure}[h]
\centering
\subfigure[]{\includegraphics[width=0.85\textwidth]{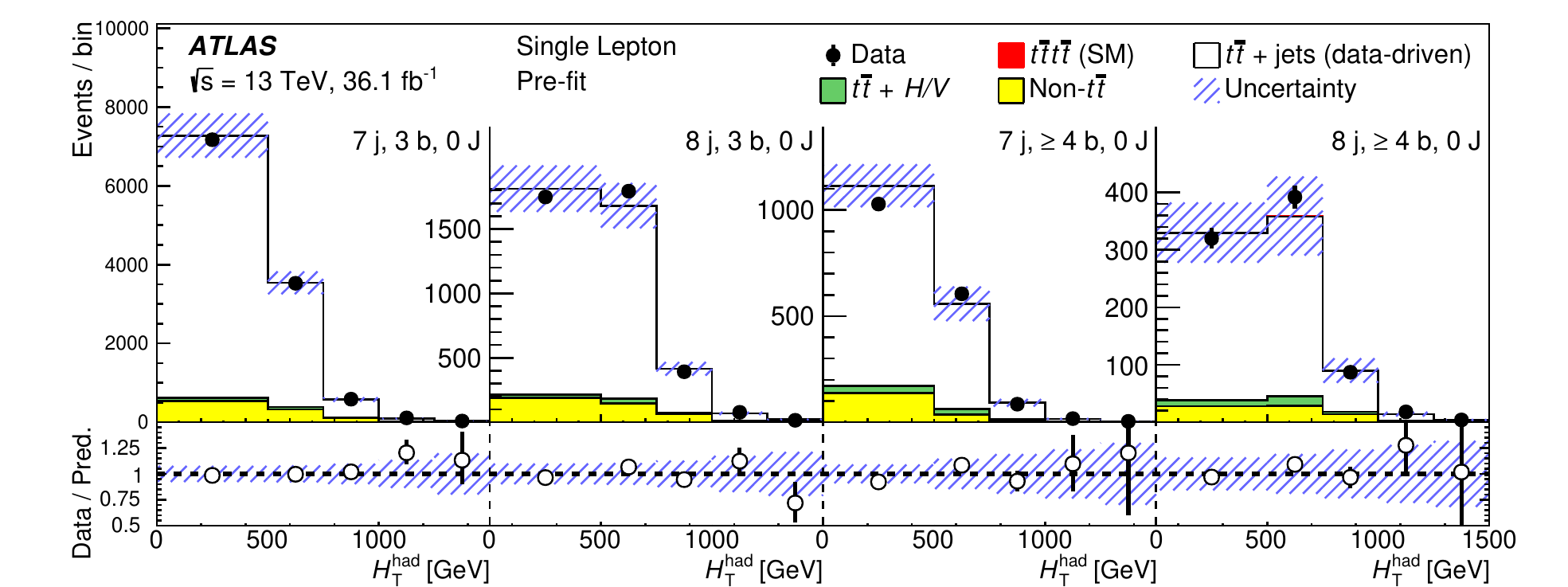}}\\
\subfigure[]{\includegraphics[width=0.85\textwidth]{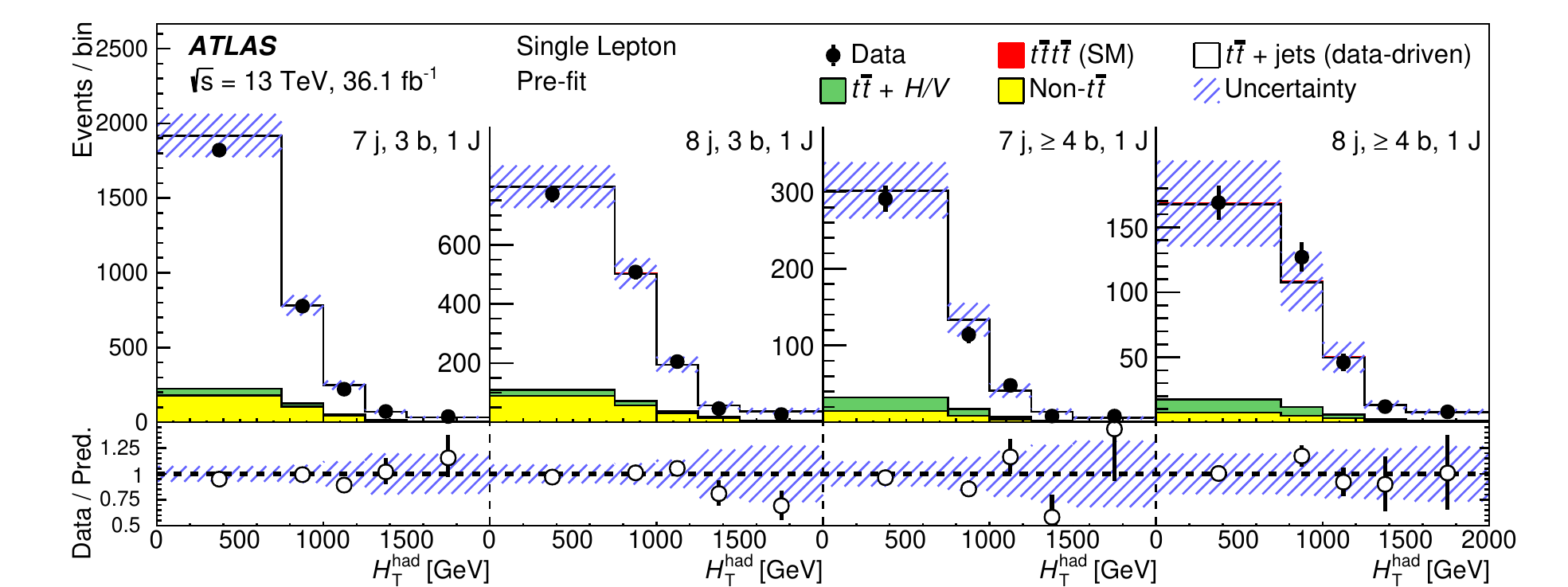}}\\
\subfigure[]{\includegraphics[width=0.85\textwidth]{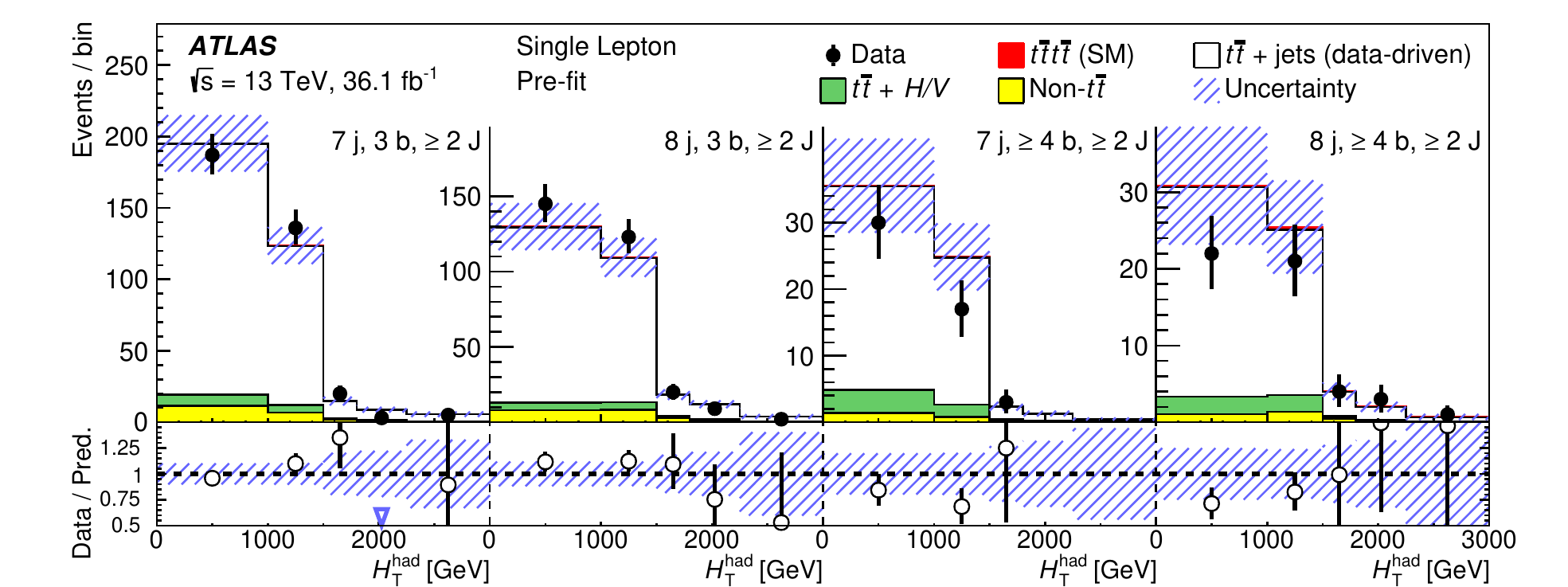}}
\caption{Comparison between data and prediction of the \hthad\ distributions in the single-lepton validation regions prior to the combined fit to data in the single-lepton and dilepton channels. The $\ttbar$+jets background is estimated with the data-driven method. The $\ttbar+V$ and $\ttbar+H$ processes are denoted $\ttbar+H/V$. Contributions from $W/Z$+jets, single-top, diboson and multijet backgrounds are combined into a single background source referred to as ``Non-$t\bar{t}$''. The hashed area represents the combined statistical and systematic uncertainties of the prediction. The last bin in all figures contains the overflow. The lower panel shows the ratio between the data and the total prediction, including the SM $t\bar{t}t\bar{t}$ signal. An arrow indicates that the point is off-scale.
}
\label{fig:validationRegions_ljets}
\end{figure}

\begin{figure}[h]
\centering
\includegraphics[width=0.85\textwidth]{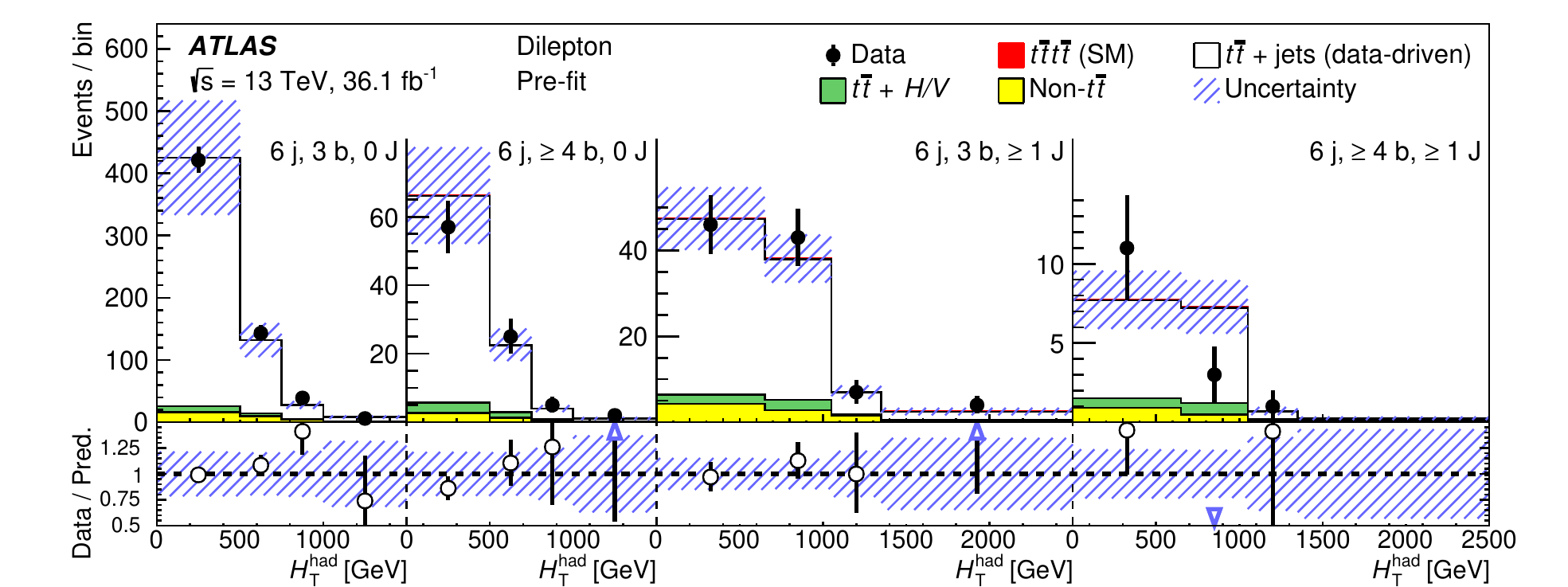}
\caption{
Comparison between data and prediction of the \hthad\ distributions in the dilepton validation regions prior to the combined fit to data in the single-lepton and dilepton channels. The $\ttbar$+jets background is estimated with the data-driven method. The $\ttbar+V$ and $\ttbar+H$ processes are denoted $\ttbar+H/V$. Contributions from $W/Z$+jets, single-top, diboson and multijet backgrounds are combined into a single background source referred to as ``Non-$t\bar{t}$''. The hashed area represents the combined statistical and systematic uncertainties of the prediction. The last bin in all figures contains the overflow. The lower panel shows the ratio between the data and the total prediction, including the SM $t\bar{t}t\bar{t}$ signal. An arrow indicates that the point is off-scale.
}
\label{fig:validationRegions_dilep}
\end{figure}

Validation regions are designed primarily to validate the TRF$_{\ttbar}$ data-driven estimate of the dominant $\ttbar$+jets background and confirm the validity of the assumption that the estimate can be extrapolated to the signal regions. Comparisons of the $\hthad$ distributions between data and the total SM prediction (including the SM four-top-quark signal) in the validation regions prior to the fit to data are presented in Figure~\ref{fig:validationRegions_ljets} for the single-lepton channel and in Figure~\ref{fig:validationRegions_dilep} for the dilepton channel. The $\ttbar$+jets background is estimated with the data-driven method, including the MC correction factors and the systematic uncertainties. Data agree well with the SM expectation within the uncertainties, validating the overall data-driven procedure and the assumptions made.

\clearpage

\section{Systematic uncertainties}
\label{sec:Systematics}
Several sources of systematic uncertainties that can affect the normalization of signal and background and the shape of the $\hthad$ distributions are considered. The systematic uncertainties of the data-driven estimate for the $\ttbar$+jets background are propagated as described in Section~\ref{sec:ttDD}. For each considered source of systematic uncertainty affecting the $\ttbar$+jets MC prediction, a new set of correction factors $C'_i$ is derived, by coherently replacing the nominal MC prediction with the systematic variation in all regions. The usage of this data-driven technique to estimate the $\ttbar$+jets background, as opposed to a purely simulation-based approach, allows to reduce significantly the uncertainty on its prediction in the high jet and $b$-tagged jet multiplicity topologies exploited by this search.

\subsection{Experimental uncertainties}
\label{sec:syst_experiment}
The uncertainty in the combined 2015+2016 integrated luminosity affecting the overall normalisation of all processes estimated from the simulation is 2.1\%. It is derived, following a methodology similar to that detailed in Ref.~\cite{Aaboud:2016hhf}, and using the LUCID-2 detector for the baseline luminosity measurements \cite{LUCID2}, from calibration of the luminosity scale using $x$--$y$ beam-separation scans. This systematic uncertainty is applied to all processes modeled using MC simulations.

Uncertainties associated with jets primarily arise from the jet energy scale. The jet energy scale (JES) and its uncertainty are derived by combining information from test-beam data, LHC collision data and simulation~\cite{newJES}. The JES uncertainty is split into 21 uncorrelated sources, which have different dependencies on jet $\pt$ and $\eta$. In particular, three uncertainties account for differences in the jet response and simulated jet composition of light-quark, $b$-quark, and gluon-initiated jets. The flavor response uncertainties are derived by comparing the average jet response for each jet flavor using {\PYTHIA} and {\Herpp}.  The flavor composition uncertainty is assumed to be a 50\% quark and 50\% gluon composition with a conservative 100\% uncertainty. Uncertainties in the jet mass scale, the jet energy resolution and the efficiency to pass the JVT requirement are also considered.

The efficiency of the $b$-tagging algorithm is measured for each jet flavor using control samples in data and in simulation. From these measurements, correction factors are derived to match the tagging rates in the simulation~\cite{Aad:2015ydr,ATLAS-CONF-2018-001,ATLAS-CONF-2018-006}. Uncertainties in these corrections include a total of six independent sources affecting $b$-jets and four independent sources affecting $c$-jets. Each uncertainty has a different dependence on jet $\pt$. Seventeen uncertainties are considered for the light-jet tagging, which depend on the jet $\pt$ and $\eta$. These systematic uncertainties are taken as uncorrelated between $b$-jets, $c$-jets, and light-flavor jets. An additional uncertainty is included due to the extrapolation of these corrections to jets with $\pt$ beyond the kinematic reach of the data calibration samples used ($\pt>300\gev$ for $b$- and $c$-jets and $\pt>750\gev$ for light-jets) and is taken to be correlated among the three jet flavors. 

Uncertainties associated with leptons arise from the trigger, reconstruction, identification, and isolation efficiencies, as well as the lepton momentum scale and resolution. These are measured in data using leptons in $Z \rightarrow \ell^{+}\ell^{-}$ and $J/\psi \rightarrow \ell^{+}\ell^{-}$ events at $\sqrt{s} = 13$~TeV~\cite{ATLAS-CONF-2016-024,muon_latest}. 

All uncertainties in energy scales and resolutions are propagated to the missing transverse momentum. Additional small uncertainties associated with the modeling of the underlying event affecting the reconstruction of the missing transverse momentum are also taken into account.

\subsection{Modeling uncertainties}
\label{sec:syst_bkgmodeling}
As mentioned in Section~\ref{sec:ttDD}, common normalization uncertainties for $\ttbar$+jets that equally affect $B^{\mathrm{MC}}_i$ and $B^{\mathrm{TRF}_{\ttbar},\mathrm{MC}}_i$ have no impact on their ratios $C_i$, and consequently on the total TRF$_{\ttbar}$ prediction. Instead, uncertainties in the $\ttbar$+jets heavy-flavor content or kinematics can have residual systematic effects on the TRF$_{\ttbar}$ prediction. Therefore, no uncertainty is assigned to the inclusive \ttbar\ production cross section in the search, while variations of the relative fractions of \ttbar\ events with additional jets originating from $b$- and $c$- quarks, as well as comparisons of $t\bar{t}$+jets kinematics with alternative predictions, are considered as systematic uncertainties related to the theory modeling of the $t\bar{t}$+jets process, as described below. A categorization of $\ttbar$+jets events is performed for the purpose of assigning systematic uncertainties associated with the modeling of heavy-flavor production in different topologies~\cite{Aad:2015gra}. Events are categorized depending on the flavor content of additional particle jets and labeled either $\ttbar$+$\geq$$1b$ or $\ttbar$+$\geq$$1c$, while the remaining events are labeled as $\ttbar$+light-jets events, including those with no additional jets.

Detailed comparisons of $\ttbar$+$\geq$$1b$ production between the nominal NLO {\POWHEGBOX} v2 + {\PYTHIA}~6.428 $\ttbar$ inclusive MC sample and an NLO prediction based on \textsc{Sherpa} + \textsc{OpenLoops}~\cite{gleisberg:2008ta, cascioli:2011va} (referred to as \textsc{SherpaOL}) have shown that the cross sections agree within 50\%~\cite{cascioli:2013era}. Therefore, a normalization uncertainty of 50\% is applied to the $\ttbar$+$\geq$$1b$ component of the $\ttbar$+jets background obtained from the {\POWHEGBOX} v2 + {\PYTHIA}~6.428 MC simulation. In the absence of an NLO prediction for the $\ttbar$+$\geq$$1c$ background, a 50\% systematic uncertainty is also applied to the $\ttbar$+$\geq$$1c$ component, and the uncertainties in the $\ttbar$+$\geq$$1b$ and $\ttbar$+$\geq$$1c$ background normalizations are taken as uncorrelated. The overall normalization of all systematic uncertainties in the $\ttbar$+jets prediction, except these explicit uncertainties in the $\ttbar$+$\geq$$1c$ and $\ttbar$+$\geq$$1b$ normalizations, is fixed to the nominal one and only migrations across categories and distortions to the shape of the kinematic distributions are considered.

To provide a comparison with a different parton-shower model, an alternative \ttbar\ sample was generated using the same {\Powheg} model setup as for the nominal sample described in Section~\ref{sec:SandB}, except the PS, hadronization, underlying-event (UE) and MPI are simulated using {\Herpp} (version 2.7.1)~\cite{bahr:2008pv} with the {\textsc UEEE5} tune~\cite{seymour:2013qka} and the corresponding CTEQ6L1 PDF set. To assess the systematic uncertainties related to the use of different models for the hard-scattering generation, while maintaining the same PS model, a sample using {\amcatnlo}~\cite{Alwall:2014hca} interfaced to {\Herpp} 2.7.1 was generated. The effects of initial- and final-state radiation (ISR/FSR) are explored using two alternative {\POWHEGBOX} v2 + {\PYTHIA}~6.428 samples, one with $h_\textrm{damp}$ set to $2 \times m_{\textrm{top}}$, the renormalization and factorization scales set to half the nominal value and using the {\textsc PERUGIA2012} high-variation UE tune, giving more radiation, and one with the {\textsc PERUGIA2012} low-variation UE tune, $h_\textrm{damp} = m_{\textrm{top}}$ and the renormalization and factorization scales set to twice the nominal value, giving less radiation~\cite{ATL-PHYS-PUB-2015-011}. The $\mu_\text{R}$ and $\mu_\text{F}$ scale variations and the $h_\textrm{damp}$ variations are kept correlated, since the two proposed variations cover the full set of uncertainties obtained by changing the scales and the resummation damping parameter independently. 

Previous studies have seen that NNLO calculations provide better agreement with data than NLO calculation, particularly for the top-quark $\pt$ distribution~\cite{Aad:2015mbv}. Hence, an uncertainty in the modeling of the top-quark $\pt$ distribution is evaluated by taking the full difference between applying and not applying the reweighting to match the predictions at NNLO accuracy in QCD~\cite{Czakon:2015owf,Czakon:2016dgf} of the top-quark $\pt$ distribution. This uncertainty only affects the $t\bar{t}$+light-jets and $\ttbar$+$\geq$$1c$ events, for which NNLO predictions have been derived in literature.

In the case of $\ttbar$+$\geq$$1b$ events, an uncertainty is assigned by comparing the NLO prediction in the four-flavor scheme of $\ttbar$+$\geq$$1b$ including parton shower~\cite{cascioli:2013era} based on \textsc{SherpaOL} with the nominal NLO {\POWHEGBOX} v2 + {\PYTHIA}~6.428 inclusive $\ttbar$ MC sample with a five-flavor scheme, by means of a generator-level reweighting, as detailed in Ref.~\cite{Aad:2015gra}. This reweighting is performed separately for each of the $\ttbar$+$\geq$$1b$ subcategories in such a way that their inter-normalization and the shape of the relevant kinematic distributions are at NLO accuracy, while preserving the nominal $\ttbar$+$\geq$$1b$ cross section in {\POWHEGBOX} v2 + {\PYTHIA}~6.428. Additional uncertainties are assessed for those contributions of $\ttbar$+$\geq$$1b$ background which are not part of the NLO prediction, namely from MPI or FSR from top-quark decay products. They are assessed via the alternative radiation samples described above.

Uncertainties affecting the modeling of the $W/Z$+jets background include 5\% scale uncertainty from their respective normalizations to the theoretical NNLO cross sections~\cite{melnikov:2006kv}. An additional 24\% normalization uncertainty is added in quadrature for each additional inclusive jet-multiplicity bin, based on a comparison among different algorithms for merging LO matrix elements and parton showers~\cite{alwall:2007fs}. Therefore, normalization uncertainties of $54\%$ and $59\%$ are assigned for events with exactly five jets and at least six jets, respectively. These normalization uncertainties are taken as correlated (uncorrelated) across jet multiplicities within signal regions (efficiency extraction regions). Uncertainties affecting the modeling of the single-top-quark background include an uncertainty of $+5\%$ and $-4\%$ in the total cross section estimated as a weighted average of the theoretical uncertainties in $t$-, $Wt$- and $s$-channel production~\cite{kidonakis:2011wy,kidonakis:2010ux,kidonakis:2010tc}. 

Uncertainties in the diboson background normalization include 5\% from the NLO cross sections~\cite{Campbell:2011bn}, as well as an additional 24\% normalization uncertainty added in quadrature for each additional inclusive jet multiplicity bin: this assumes that two of the jets originate from the $W/Z$ decays, as in $WW/WZ \to \ell \nu jj$. Recent comparisons between data and \textsc{Sherpa}~2.1.1 for $WZ(\to \ell\nu\ell\ell) + \geq$4 jets show agreement within the experimental uncertainty of approximately 40\%~\cite{Aaboud:2016yus}, which further justifies the above uncertainties. Uncertainties in the $\ttbar+V$ and $\ttbar+H$ normalizations are $\pm 15\%$ and $^{+10}_{-13}\%$, respectively, from the uncertainties in their respective NLO cross sections~\cite{campbell:2012dh,garzelli:2012bn,dittmaier:2011ti,dittmaier:2012vm}. 

For the determination of the SM \tttt\ production signal strength, no uncertainty is assigned to the theoretical cross section for this process. In the extraction of the exclusion limits on \tttt\ production via BSM models, the SM \tttt\ process is considered as background and a conservative 50\% normalization uncertainty is assigned to the total \tttt\ cross section, taking into account the uncertainties in both its production cross section and possible acceptance and shape variations~\cite{bevilacqua:2012constraining, Alwall:2014hca}.

Uncertainties in the data-driven fake or non-prompt lepton background estimate include contributions from the limited sample size in data, particularly at high jet and $b$-tag multiplicities, from the uncertainty in the real and fake efficiencies extracted from data in dedicated control regions (e.g.~selected with a requirement on either the maximum $\met$ or $\mtw$), as well as from the extrapolation from these control regions to the analysis regions, as detailed in Ref.~\cite{atlas-conf-2014-058}. Based on comparisons between data and the total prediction in these control regions, the normalization uncertainties assumed for this background are 50\% (100\%) for events with a central (forward) electron, and 50\% for muons, taken to be uncorrelated across regions with different \mreclust\ jet multiplicities and between electron and muon channels. No explicit shape uncertainty is assigned due to the associated large statistical uncertainties. These uncertainties are uncorrelated between bins in the final discriminant distribution and effectively cover possible shape uncertainties.

\section{Results}
\label{sec:Results}
Following the statistical method presented below, four-top-quark production signals are searched for by performing a binned profile likelihood fit to the $\hthad$ distribution simultaneously in the 12 signal regions in the single-lepton channel and 8 signal regions in the dilepton channel, using a total of 20 final-state topologies. The single-lepton and dilepton channels are combined in order to gain sensitivity to different four-top-quark production signals.

\subsection{Statistical interpretation}
\label{sec:StatMethods}
For each search, the $\hthad$ distributions across all regions considered are jointly analyzed to test for the presence of a signal predicted by the benchmark scenarios. The statistical interpretation uses a binned likelihood function ${\cal L}(\mu,\boldsymbol{\theta})$ constructed as a product of Poisson probability terms over all bins considered in each search (namely, all \hthad\ bins in the 20 signal regions defined in Figure~\ref{fig:ttDD_regions}). The likelihood function depends on the signal-strength parameter $\mu$, a multiplicative factor that scales the number of expected signal events, and $\boldsymbol{\theta}$, a set of nuisance parameters (NPs) that encode the effect of systematic uncertainties on the signal and background expectations, which are implemented in the likelihood function as Gaussian, log-normal or Poisson constraints. Individual sources of systematic uncertainty are considered to be uncorrelated. Correlations of a given systematic uncertainty are maintained across processes and channels. The statistical uncertainty of the prediction, which incorporates the statistical uncertainty of the MC events and of the data-driven fake and non-prompt lepton estimate, is included in the likelihood in the form of additional nuisance parameters, one for each of the included bins.

The test statistic $q_\mu$ is defined as the profile likelihood ratio: $q_\mu = -2\ln({\cal L}(\mu,\boldsymbol{\hat{\hat{\theta}}}_\mu)/{\cal L}(\hat{\mu},\boldsymbol{\hat{\theta}}))$, where $\hat{\mu}$ and $\boldsymbol{\hat{\theta}}$ are the values of the parameters that maximize the likelihood function (with the constraint $0\leq \hat{\mu} \leq \mu$), and $\boldsymbol{\hat{\hat{\theta}}}_\mu$ are the values of the NPs that maximize the likelihood function for a given value of $\mu$. The test statistic $q_\mu$ is implemented in the \textsc{RooFit} package~\cite{verkerke:2003ir,RooFitManual}. In the absence of any significant excess above the background expectation, upper limits on the signal production cross section for each of the signal scenarios considered in Section~\ref{sec:Signal} are derived by using $q_\mu$ and the CL$_{\textrm{s}}$ method~\cite{junk:1999kv,read:2002hq}. For a given signal scenario, values of the production cross section (parameterized by $\mu$) yielding CL$_{\textrm{s}}$$<0.05$, where CL$_{\textrm{s}}$ is computed using the asymptotic approximation~\cite{cowan:2010js}, are excluded at >95\% C.L.

\subsection{Comparison between data and prediction in signal regions after the fit to data}
\label{sec:finalAn_fullunblinding}
\begin{figure}[ht!]
\centering
\subfigure[]{\includegraphics[width=0.85\textwidth]{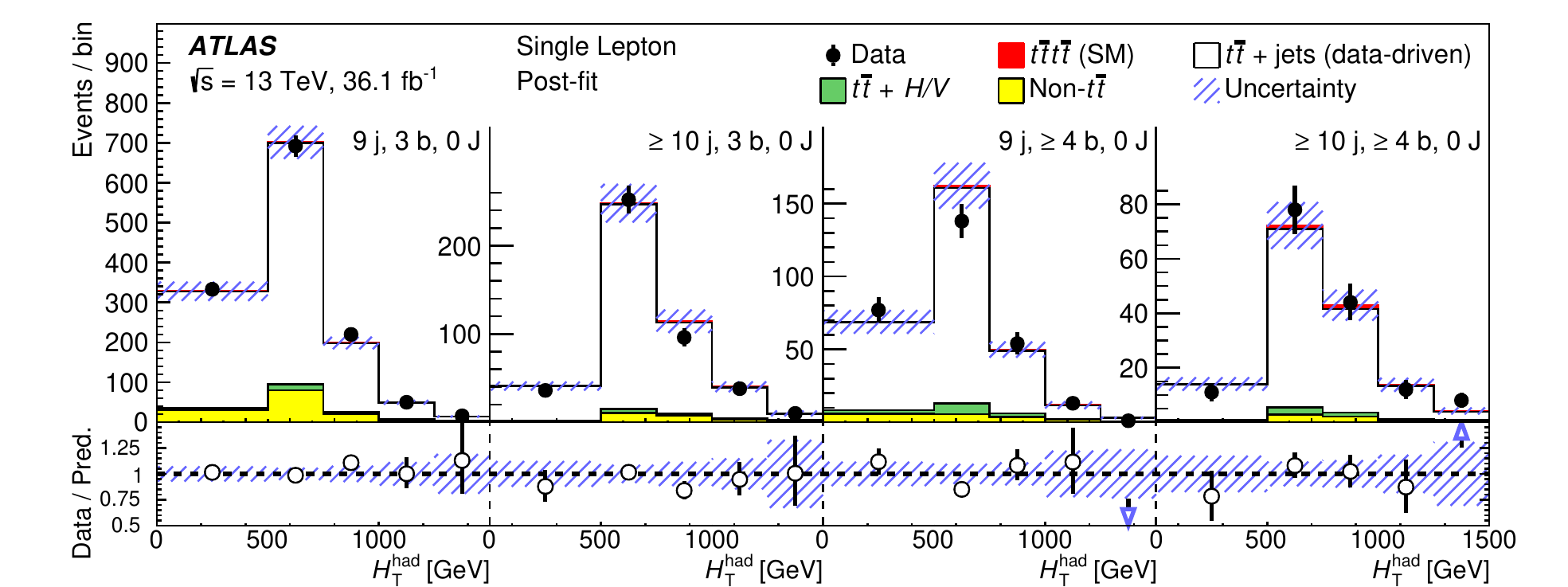}}\\
\subfigure[]{\includegraphics[width=0.85\textwidth]{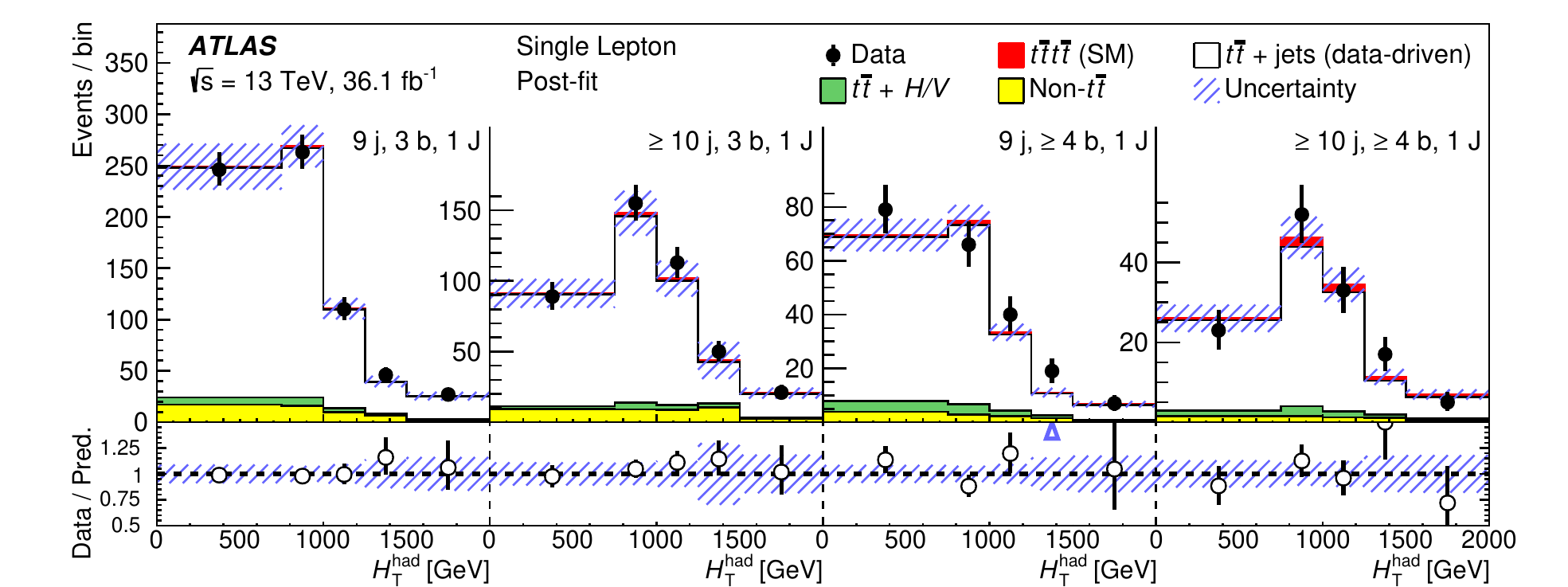}}\\
\subfigure[]{\includegraphics[width=0.85\textwidth]{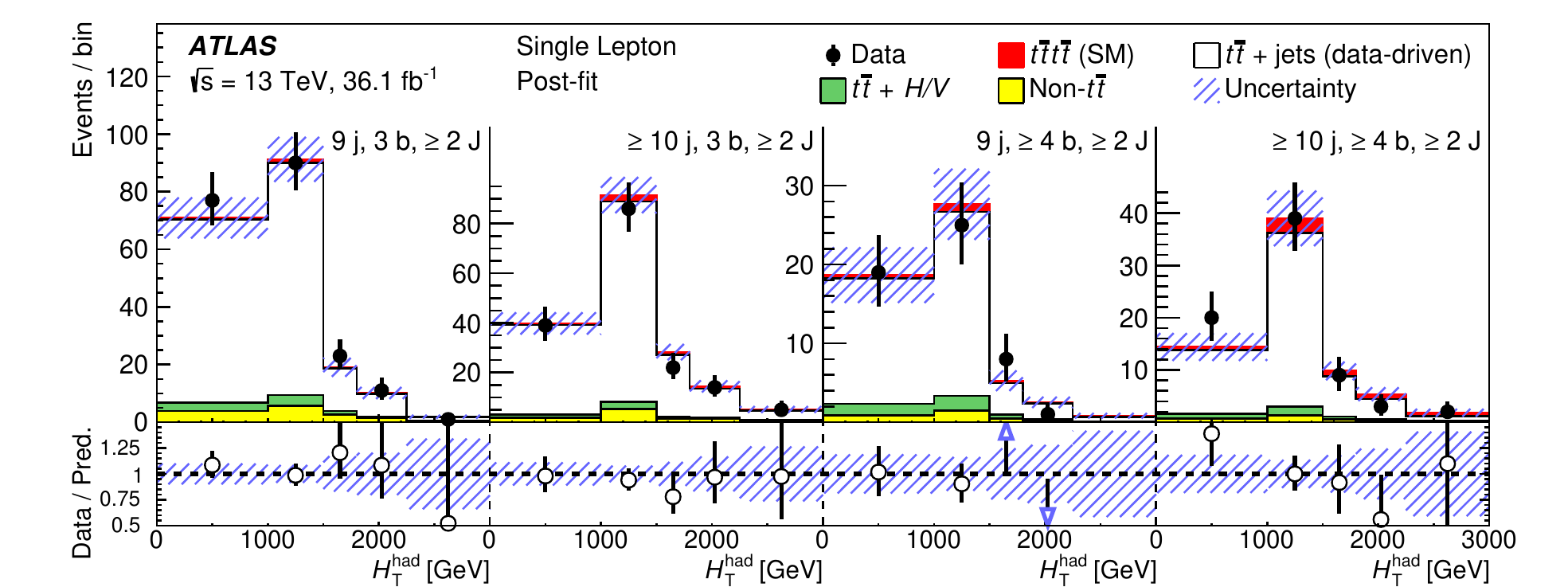}}
\caption{
Comparison between data and prediction of the \hthad\ distributions in the single-lepton signal regions after the combined fit to data in both the single-lepton and dilepton channels. The $\ttbar$+jets background is estimated with the data-driven method. The $\ttbar+V$ and $\ttbar+H$ processes are denoted $\ttbar+H/V$. Contributions from $W/Z$+jets, single-top, diboson and multijet backgrounds are combined into a single background source referred to as ``Non-$t\bar{t}$''. The hashed area represents the combined statistical and systematic uncertainties of the prediction. The last bin in all figures contains the overflow. The lower panel shows the ratio between the data and the total prediction, including the SM \tttt\ signal scaled by the best-fit signal strength. An arrow indicates that the point is off-scale.
}
\label{fig:signalRegions_comb_ljets}
\end{figure}

\begin{figure}[ht!]
\centering
\subfigure[]{\includegraphics[width=0.85\textwidth]{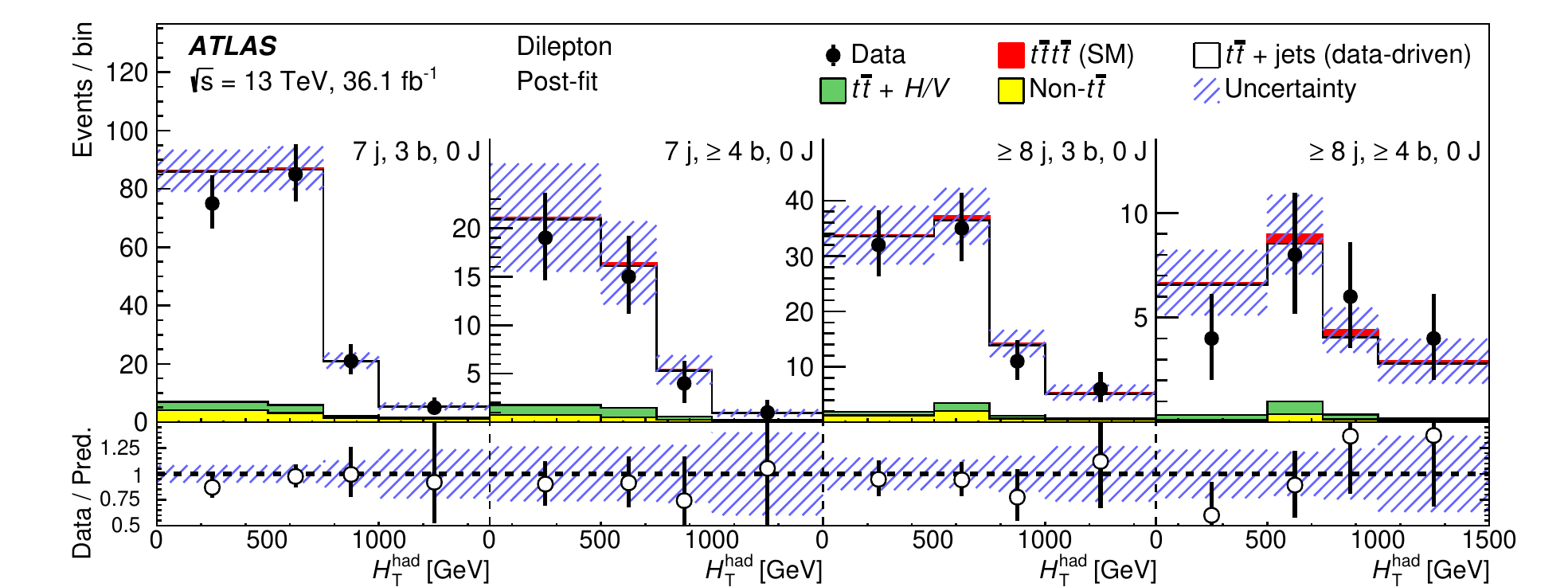}}\\
\subfigure[]{\includegraphics[width=0.85\textwidth]{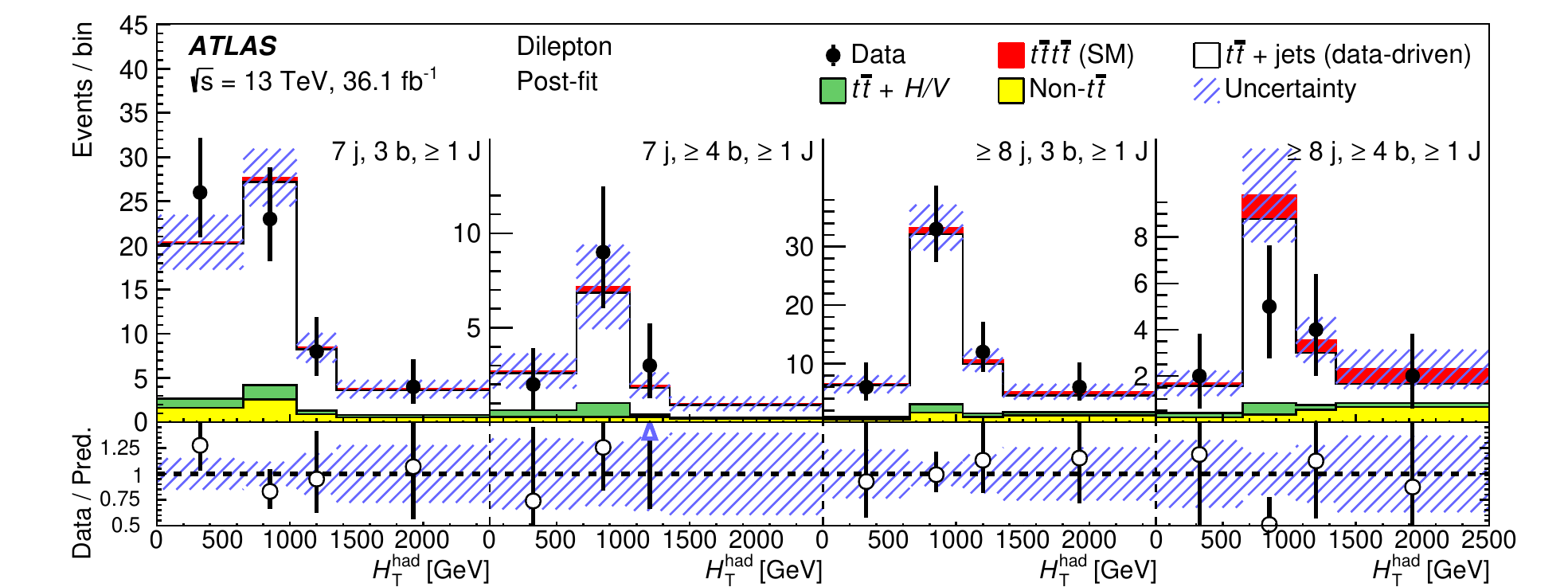}}
\caption{
Comparison between data and prediction of the \hthad\ distributions in the dilepton signal regions after the combined fit to data in both the single-lepton and dilepton channels. The $\ttbar$+jets background is estimated with the data-driven method. The $\ttbar+V$ and $\ttbar+H$ processes are denoted $\ttbar+H/V$. Contributions from $W/Z$+jets, single-top, diboson and multijet backgrounds are combined into a single background source referred to as ``Non-$t\bar{t}$''. The hashed area represents the combined statistical and systematic uncertainties of the prediction. The last bin in all figures contains the overflow. The lower panel shows the ratio between the data and the total prediction, including the SM \tttt\ signal scaled by the best-fit signal strength. An arrow indicates that the point is off-scale.
}
\label{fig:signalRegions_comb_dilep}
\end{figure}

A binned likelihood fit to the data is performed in the 12 signal regions in the single-lepton channel and 8 signal regions in the dilepton channel, leading to good agreement between data and post-fit estimates. Comparisons of the $\hthad$ distributions between data and the total SM prediction (including the SM \tttt\ signal) in the signal regions, after the combined fit to data in the signal-plus-background hypothesis in the two channels, are presented in Figure~\ref{fig:signalRegions_comb_ljets} for the single-lepton channel and in Figure~\ref{fig:signalRegions_comb_dilep} for the dilepton channel. Good agreement of the extrapolated fit results is observed as well in the validation regions, which are presented in Appendix~\ref{app:VRpost-fit}. 

\pagebreak
\clearpage

Table~\ref{tab:syst_impact} shows the post-fit impact of the largest sources of systematic uncertainty on the signal strength $\mu$ after the simultaneous fit to data in the single-lepton and dilepton channels. The leading sources of systematic uncertainty vary depending on the analysis region considered. The largest contributions are due to the uncertainty associated with the choice of $\ttbar$+jets parton shower and hadronization model and that of the $\ttbar$+jets NLO generator, as well as large statistical uncertainties associated with the background prediction.

\begin{table}
\sisetup{table-format=1.3,retain-explicit-plus}
\centering
\caption{
Breakdown of the contributions to the uncertainties on $\mu$. The quoted uncertainties $\Delta\mu$ are obtained by repeating the fit with certain sets of nuisance parameters fixed to their post-fit values, and subtracting in quadrature the resulting total uncertainty of $\mu$ from the uncertainty from the full fit. The total statistical uncertainty is evaluated by fixing all nuisance parameters in the fit. The line ``background-model statistical uncertainty'' refers to the statistical uncertainties of the MC event samples and in the data-driven determination of the $\ttbar$+jets and the non-prompt and fake-lepton background components. These uncertainties are evaluated after the fit described in Section~\ref{sec:Results}. 
}
\begin{tabular}{l S S}
\toprule\toprule
Uncertainty source          & \multicolumn{2}{c}{$\pm \Delta\mu $}\\
\midrule\midrule
$\ttbar$+jets modeling                    & +1.2    & -0.96 \\
Background-model statistical uncertainty  & +0.91   & -0.85 \\
Jet energy scale and resolution, jet mass & +0.38   & -0.16 \\
Other background modeling                 & +0.26   & -0.20 \\
$b$-tagging efficiency and mis-tag rates  & +0.33   & -0.10 \\
JVT, pileup modeling                      & +0.18   & -0.073 \\
$\ttbar + H/V$ modeling                   & +0.053  & -0.055 \\
Luminosity                                & +0.050  & -0.026 \\
\midrule
Total systematic uncertainty              & +1.6    & -1.4 \\
\midrule
Total statistical uncertainty             & +1.1    & -1.0 \\
\midrule
Total uncertainty                         & +1.9    & -1.7 \\
\bottomrule\bottomrule
\end{tabular}
\label{tab:syst_impact}
\end{table}

\subsection{Limits on four-top-quark production in the single-lepton and dilepton channel}
\label{sec:Results_us}
No significant excess of events above the SM background prediction, excluding the SM $\tttt$ production, is found. In the case of $\tttt$ production with SM kinematics, an observed (expected) 95\% C.L. upper limit on the production cross section of 47~fb (33~fb) is obtained, corresponding to an upper limit on $\sigma(\tttt)$ relative to the SM prediction of 5.1~(3.6). The SM fitted signal strength $\mu$, after combination of the single-lepton and dilepton channels, is measured to be $1.7^{+1.9}_{-1.7}$. 

The search is used to set limits on BSM four-top-quark production via an EFT model (see Section~\ref{sec:SandB}). For setting limits on this BSM model, the SM \tttt\ process is considered as a background. In the case of $\tttt$ production via an EFT model with a four-top-quark contact interaction, an observed (expected) 95\% C.L. upper limit on the production cross section of 21~fb (22~fb) is obtained. The cross-section limit for the contact interaction case is lower than in the SM because the contact interaction tends to result in final-state objects with slightly larger momenta (see e.g. Figure~\ref{fig:sep2}). The upper limit on the production cross section can be translated into an observed (expected) limit on the free parameter of the model $|C_{4t}| / \Lambda^2  < 1.9$~TeV$^{-2}$~(1.9~TeV$^{-2}$).

\subsection{Combination with the same-sign dilepton and multilepton final-state search}
\label{sec:combo}
The ATLAS Collaboration has carried out a search for new physics using $36.1$\ fb$^{-1}$ of $pp$ collisions at $\sqrt{s} = 13$ TeV in the same-sign dilepton and multilepton final states (referred to as ``SS dilepton / trilepton'' channel)~\cite{SS_4tops}. In order to improve the sensitivity to final states containing four top quarks, the results of the search in single-lepton events or dilepton events with two opposite-sign charged leptons reported in Section~\ref{sec:Results_us} (referred to as ``single lepton / OS dilepton'' channel) are combined with the results from the complementary SS dilepton / trilepton channel. 

In the combination, all the experimental systematic uncertainties (described in Section~\ref{sec:syst_experiment}) are treated as fully correlated between the two channels, while all the background modeling systematic uncertainties described in Section~\ref{sec:syst_bkgmodeling} are kept uncorrelated with those in the SS dilepton / trilepton channel. This choice is motivated by the different nature of most of the background contributions in the two channels, the different importance of the common background processes and the different techniques used for the data-driven estimates.

The expected sensitivity to the SM \tttt\ production from the combination of the two searches, expressed in terms of signal significance relative to the background-only prediction, is 1.0 standard deviation, while the observed value is 2.8 standard deviations. The excess is driven by the SS dilepton / trilepton channel, where the observed (expected) SM \tttt\ signal significance amounts to 3.0 (0.8) standard deviations, to be compared with the 1.0 (0.6) standard deviation found in the single lepton / OS dilepton search. The kinematic properties of the SS dilepton / trilepton events were compared with the expectations from the BSM \tttt\ production benchmark models studied therein, and found to agree poorly with all of them, in particular for the $b$-tagged jet multiplicity.

\begin{figure}[ht!]
\centering
\subfigure[]{\includegraphics[width=0.49\textwidth]{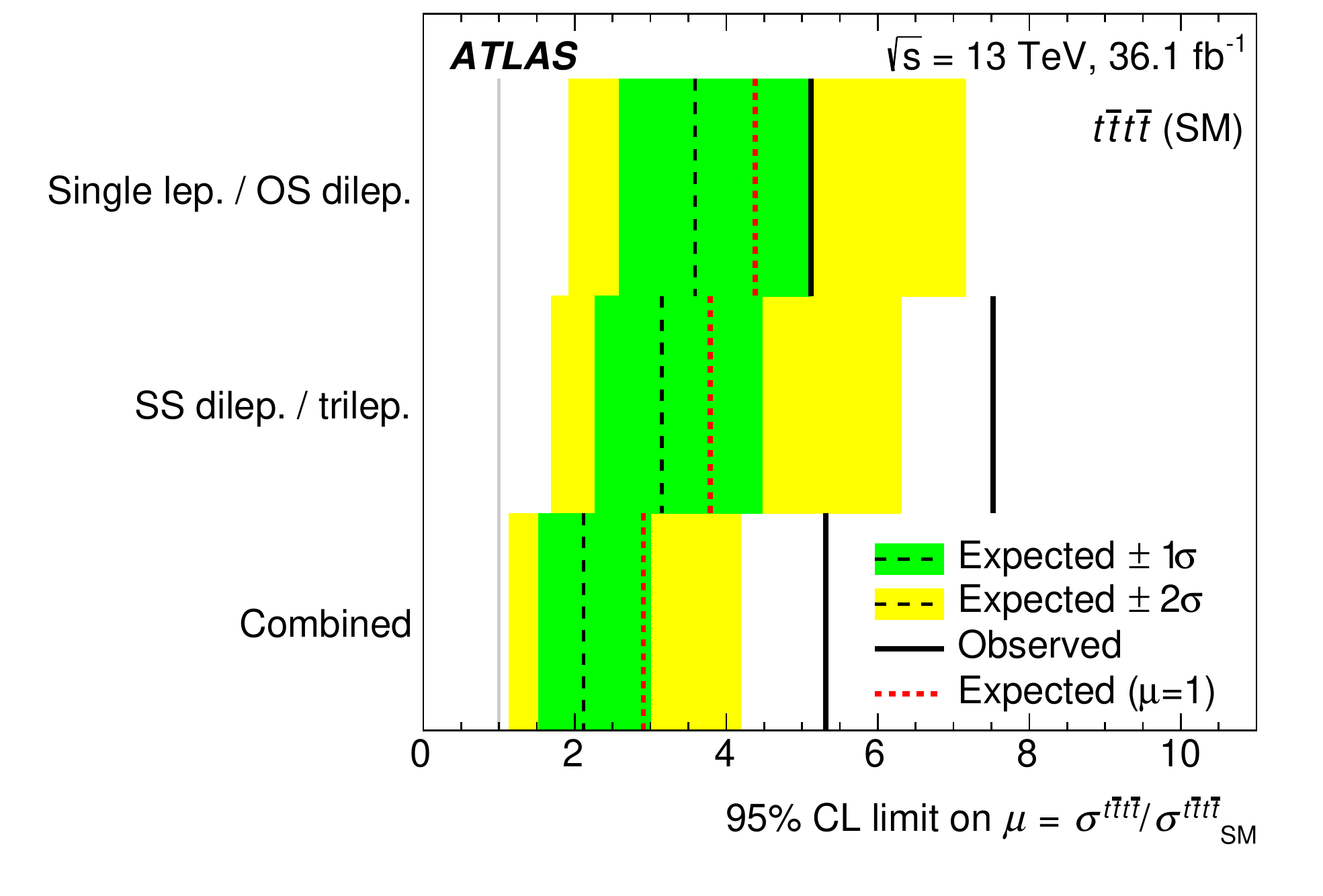}}
\subfigure[]{\includegraphics[width=0.49\textwidth]{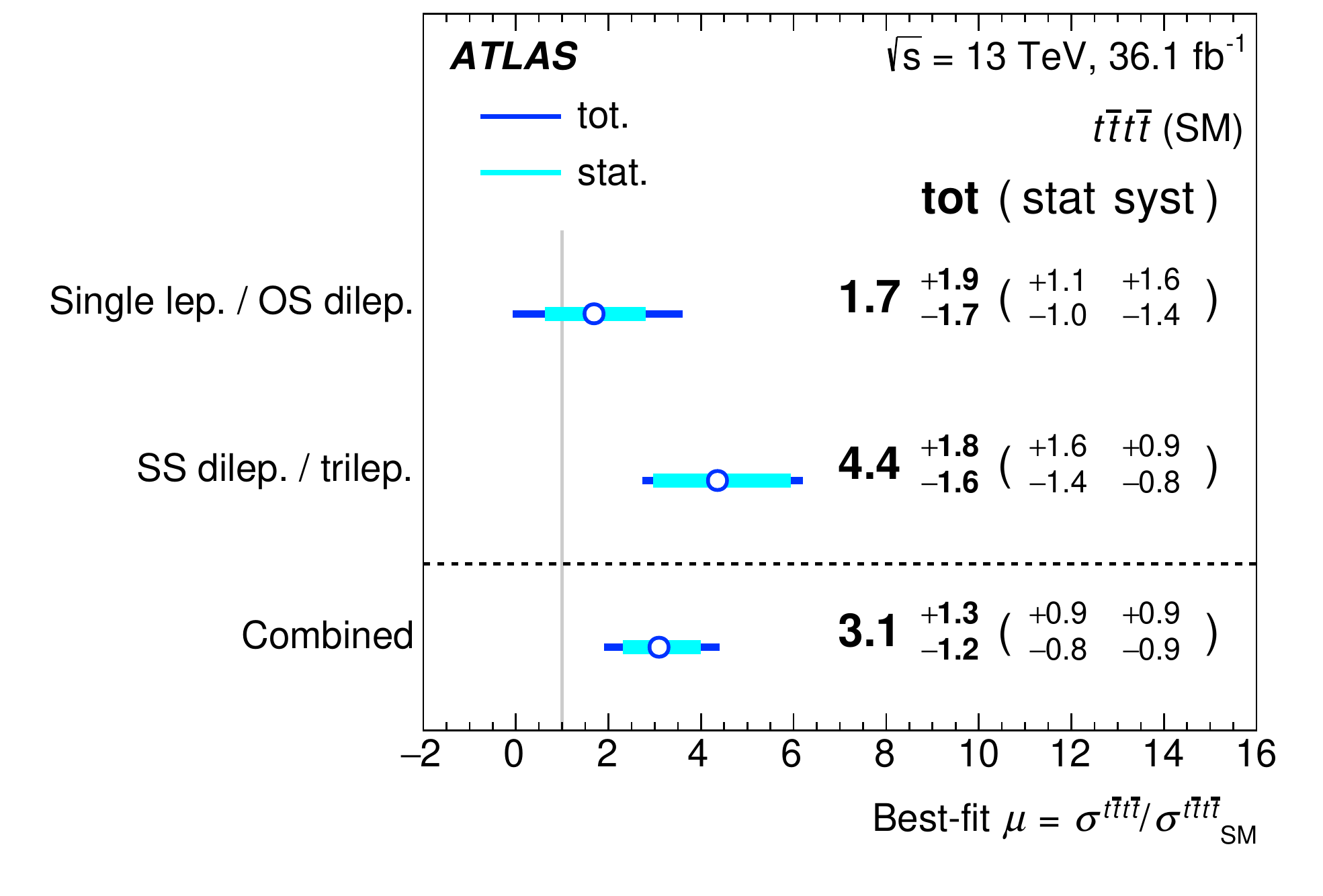}}
\caption{(a) Summary of the 95\% C.L. upper limits on $\sigma($\tttt$)$ relative to the SM prediction in the individual channels and for the combination. The observed limits (solid black lines) are shown together with the expected limits in the background-only hypothesis (dashed black lines) and in the SM signal-plus-background hypothesis case (dashed red lines). One- and two-standard-deviation uncertainty bands around the expected limits in the background-only hypothesis are also shown. (b) Summary of the signal-strength measurements in the individual channels and for the combination. The statistical uncertainties are evaluated from a fit to the data performed with all the nuisance parameters associated with systematic uncertainties fixed to their post-fit values from the nominal fit.
}
\label{fig:Limits_combo}
\end{figure}

Assuming no signal, an observed (expected) 95\% C.L. upper limit on the SM four-top-quark production cross section of 49~fb (19~fb) is obtained. It corresponds to an upper limit on $\sigma(\tttt)$ relative to the SM prediction of 5.3~(2.1). In the signal-plus-background hypothesis, the best-fit value of the SM cross section is found to be $\sigma_{\mathrm{SM}}^{\tttt} = 28.5^{+12}_{-11}$ fb, to be compared to the theoretical prediction of $9.2^{+2.9}_{-2.4}$~(scale)~$\pm 0.5$~(PDF)~fb~\cite{Alwall:2014hca}. Figure~\ref{fig:Limits_combo}(a) shows the expected and observed upper limits on $\sigma_{\mathrm{SM}}^{\tttt}$ for the two searches separately and for the combined search, while Figure~\ref{fig:Limits_combo}(b) shows a summary of the signal-strength measurements for each of the two searches and their combination. In the SS dilepton / trilepton channel the uncertainty in $\mu$ is mainly statistical, while the systematic uncertainties dominate the sensitivity of the search in the single lepton / OS dilepton channel. The probability that the results of the two searches are compatible is assessed by comparing the maximum-likelihood values for a fit performed after decorrelating the signal-strength parameters in the two channels and for the nominal combined fit with a common signal-strength parameter. The probability of obtaining a discrepancy between the two signal-strength parameters equal to or larger than the one obtained is found to be 31\%.

Limits are also set for BSM $\tttt$ production via an EFT model with a four-top-quark contact interaction. In this benchmark scenario, the SM \tttt\ sample is included as a  background process. A combined observed (expected) 95\% C.L. upper limit on the production cross section of 21~fb (15~fb) is obtained, which translates into an observed (expected) limit on the free parameter of the model of $|C_{4t}| / \Lambda^2  < 1.9$~TeV$^{-2}$~(1.6~TeV$^{-2}$).

\section{Summary}
\label{sec:Summary}
A search for four-top-quark production in the single-lepton and opposite-sign dilepton channels is presented. The analyzed data sample consists of \totlumi\ of proton--proton collisions at $\sqrt{s} = 13$~\TeV\, collected with the ATLAS detector at the Large Hadron Collider during 2015 and 2016. In order to improve the sensitivity of the search, events are categorized according to their jet, $b$-tagged jet and mass-tagged reclustered large-$R$ jet multiplicities. No significant excess of events above the SM background expectation is found. For the four-top-quark Standard Model production, an observed (expected) 95\% C.L. upper limit on the production cross section of 47~fb (33~fb), corresponding to 5.1~(3.6) times the SM prediction, is obtained. The result is combined with the same-sign dilepton and multilepton final-states search carried out by ATLAS~\cite{SS_4tops} and an observed (expected) upper limit of 49 fb (19 fb), corresponding to 5.3 (2.1) times the SM prediction is obtained at 95\% C.L. Additionally, in the case of four-top-quark production via an EFT model with a four-top-quark contact interaction, a combined observed (expected) 95\% C.L. upper limit on the production cross section of 21~fb (15~fb) is obtained.

\section*{Acknowledgments}
We thank CERN for the very successful operation of the LHC, as well as the
support staff from our institutions without whom ATLAS could not be
operated efficiently.

We acknowledge the support of ANPCyT, Argentina; YerPhI, Armenia; ARC, Australia; BMWFW and FWF, Austria; ANAS, Azerbaijan; SSTC, Belarus; CNPq and FAPESP, Brazil; NSERC, NRC and CFI, Canada; CERN; CONICYT, Chile; CAS, MOST and NSFC, China; COLCIENCIAS, Colombia; MSMT CR, MPO CR and VSC CR, Czech Republic; DNRF and DNSRC, Denmark; IN2P3-CNRS, CEA-DRF/IRFU, France; SRNSFG, Georgia; BMBF, HGF, and MPG, Germany; GSRT, Greece; RGC, Hong Kong SAR, China; ISF and Benoziyo Center, Israel; INFN, Italy; MEXT and JSPS, Japan; CNRST, Morocco; NWO, Netherlands; RCN, Norway; MNiSW and NCN, Poland; FCT, Portugal; MNE/IFA, Romania; MES of Russia and NRC KI, Russian Federation; JINR; MESTD, Serbia; MSSR, Slovakia; ARRS and MIZ\v{S}, Slovenia; DST/NRF, South Africa; MINECO, Spain; SRC and Wallenberg Foundation, Sweden; SERI, SNSF and Cantons of Bern and Geneva, Switzerland; MOST, Taiwan; TAEK, Turkey; STFC, United Kingdom; DOE and NSF, United States of America. In addition, individual groups and members have received support from BCKDF, CANARIE, CRC and Compute Canada, Canada; COST, ERC, ERDF, Horizon 2020, and Marie Sk{\l}odowska-Curie Actions, European Union; Investissements d' Avenir Labex and Idex, ANR, France; DFG and AvH Foundation, Germany; Herakleitos, Thales and Aristeia programmes co-financed by EU-ESF and the Greek NSRF, Greece; BSF-NSF and GIF, Israel; CERCA Programme Generalitat de Catalunya, Spain; The Royal Society and Leverhulme Trust, United Kingdom. 

The crucial computing support from all WLCG partners is acknowledged gratefully, in particular from CERN, the ATLAS Tier-1 facilities at TRIUMF (Canada), NDGF (Denmark, Norway, Sweden), CC-IN2P3 (France), KIT/GridKA (Germany), INFN-CNAF (Italy), NL-T1 (Netherlands), PIC (Spain), ASGC (Taiwan), RAL (UK) and BNL (USA), the Tier-2 facilities worldwide and large non-WLCG resource providers. Major contributors of computing resources are listed in Ref.~\cite{ATL-GEN-PUB-2016-002}.

\clearpage

\appendix
\part*{Appendix}
\addcontentsline{toc}{part}{Appendix}

\section{Comparison between data and prediction of the \hthad\ distributions in validation regions after the fit to data} 
\label{app:VRpost-fit}
\begin{figure}[ht!]
\centering
\includegraphics[width=0.75\textwidth]{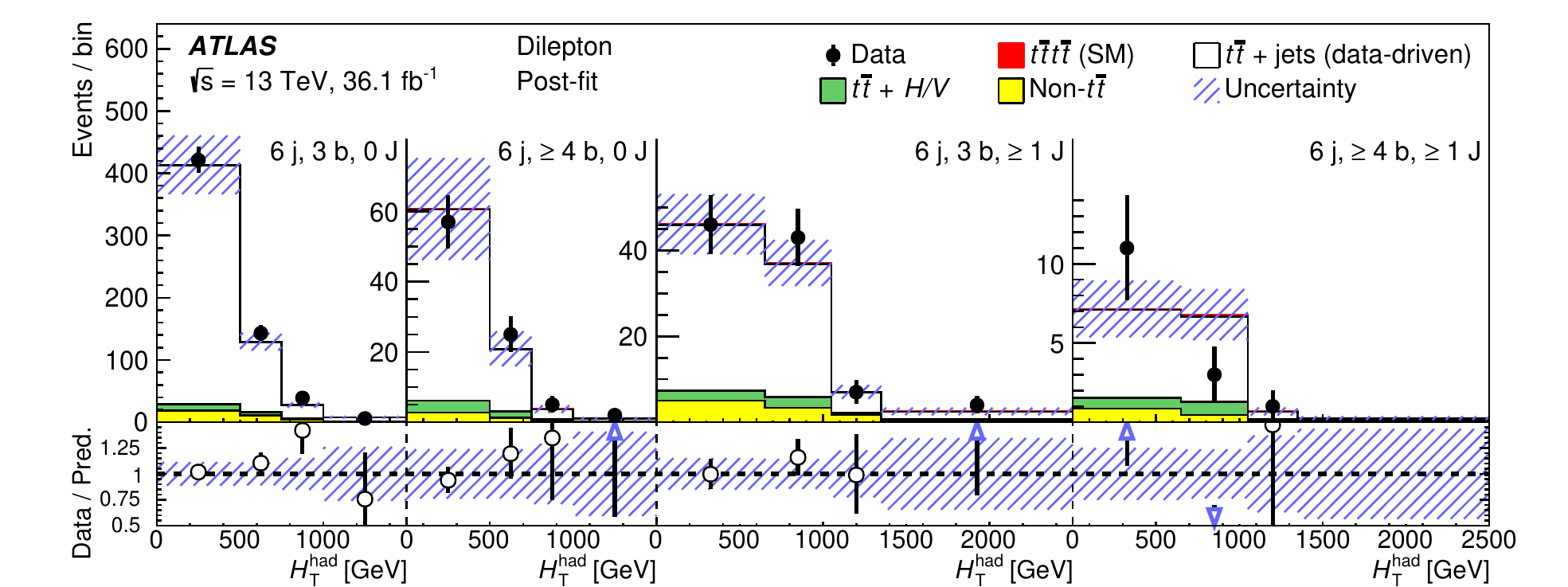}
\caption{
Comparison between data and prediction of the \hthad\ distributions in the dilepton validation regions after the combined fit to data in the single-lepton and dilepton channels. The $\ttbar$+jets background is estimated with the data-driven method. The $\ttbar+V$ and $\ttbar+H$ processes are denoted $\ttbar+H/V$. Contributions from $W/Z$+jets, single-top, diboson and multijet backgrounds are combined into a single background source referred to as ``Non-$t\bar{t}$''. The hashed area represents the combined statistical and systematic uncertainties of the prediction. The last bin in all figures contains the overflow. The lower panel shows the ratio between the data and the total prediction, including the SM $t\bar{t}t\bar{t}$ signal scaled by the best-fit signal strength. An arrow indicates that the point is off-scale.
}
\label{fig:validationRegions_post_dilep}
\end{figure}

\begin{figure}[ht!]
\centering
\subfigure[]{\includegraphics[width=0.75\textwidth]{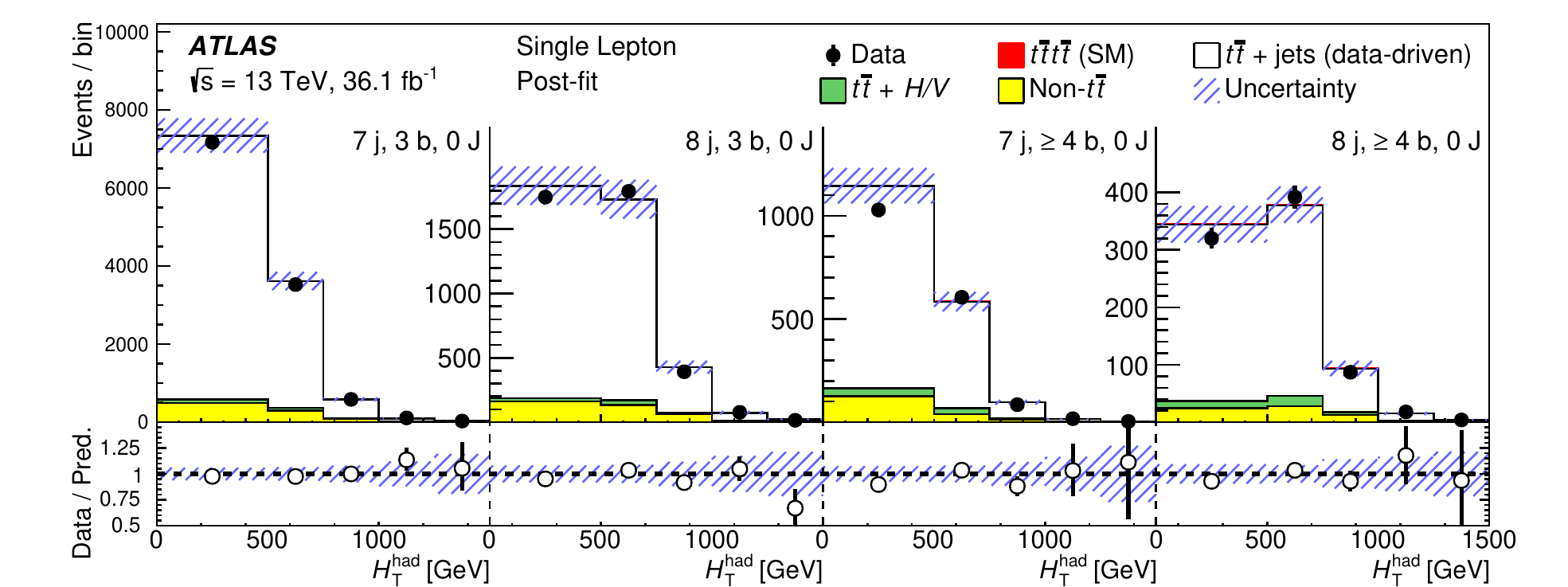}}\\
\subfigure[]{\includegraphics[width=0.75\textwidth]{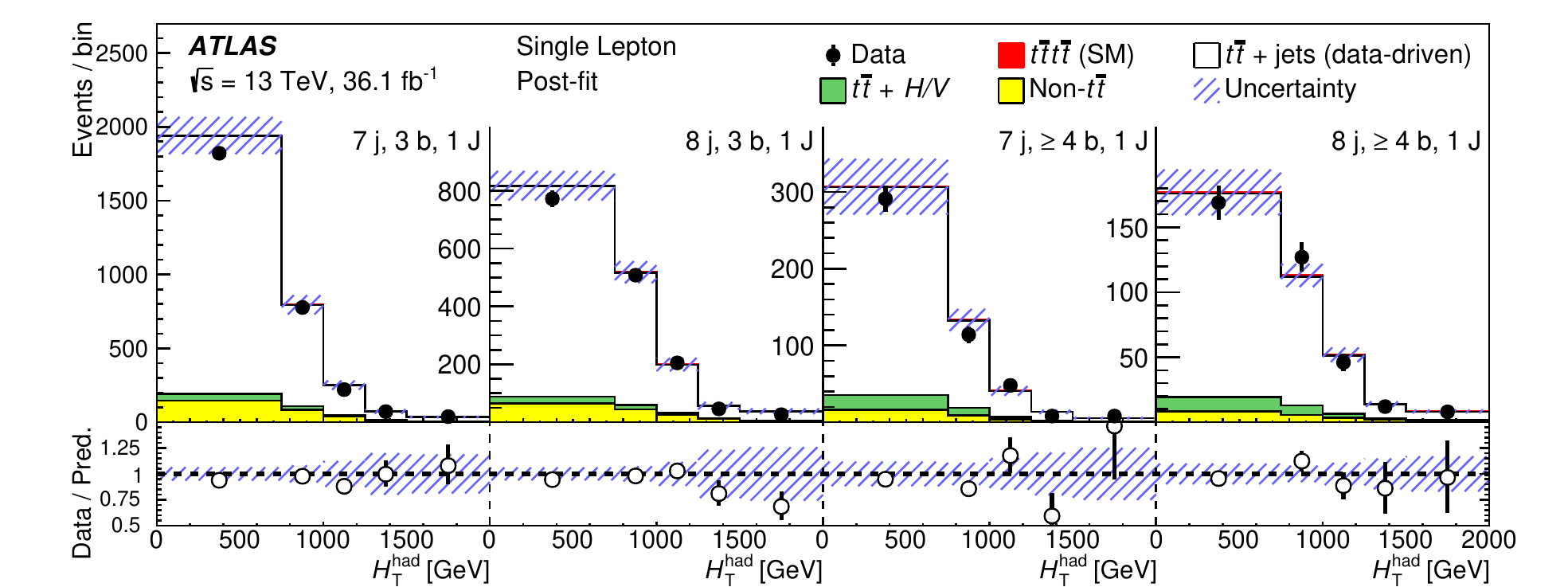}}\\
\subfigure[]{\includegraphics[width=0.75\textwidth]{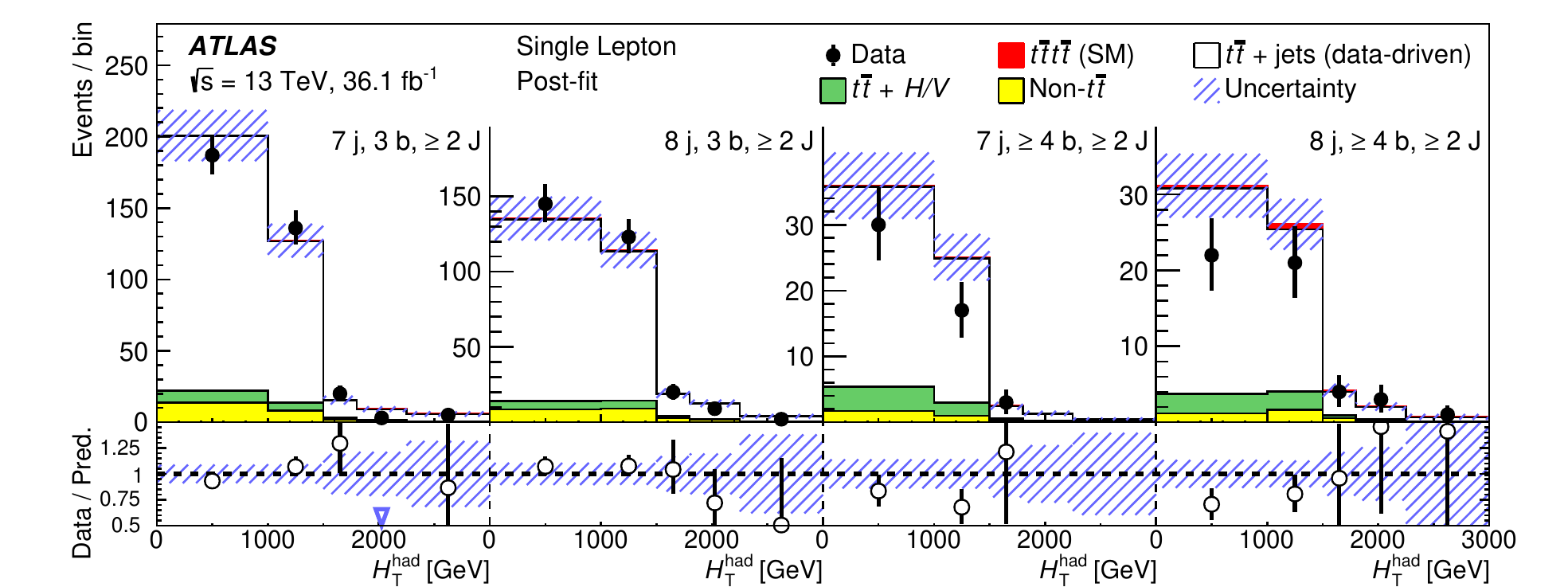}}
\caption{
Comparison between data and prediction of the \hthad\ distributions in the single-lepton validation regions after the combined fit to data in the single-lepton and dilepton channels. The $\ttbar$+jets background is estimated with the data-driven method. The $\ttbar+V$ and $\ttbar+H$ processes are denoted $\ttbar+H/V$. Contributions from $W/Z$+jets, single-top, diboson and multijet backgrounds are combined into a single background source referred to as ``Non-$t\bar{t}$''. The hashed area represents the combined statistical and systematic uncertainties of the prediction. The last bin in all figures contains the overflow. The lower panel shows the ratio between the data and the total prediction, including the SM $t\bar{t}t\bar{t}$ signal scaled by the best-fit signal strength. An arrow indicates that the point is off-scale.
}
\label{fig:validationRegions_post_ljets}
\end{figure}

Figures~\ref{fig:validationRegions_post_dilep} and~\ref{fig:validationRegions_post_ljets} show comparisons of the \hthad\ distributions in the validation regions between the data and the post-fit prediction in the dilepton and in the single-lepton channels, respectively. The post-fit prediction is obtained from the fit to data presented in Section~\ref{sec:Results} in the 20 signal regions and propagated to the validation regions, which are not included in the fit nor used to extract information from the data. The good level of agreement found between data and prediction in these regions is therefore an indication of the validity of the extrapolation of the results of the fit between different regions.

\clearpage
\printbibliography

\clearpage
 
\begin{flushleft}
{\Large The ATLAS Collaboration}

\bigskip

M.~Aaboud$^\textrm{\scriptsize 34d}$,    
G.~Aad$^\textrm{\scriptsize 99}$,    
B.~Abbott$^\textrm{\scriptsize 125}$,    
O.~Abdinov$^\textrm{\scriptsize 13,*}$,    
B.~Abeloos$^\textrm{\scriptsize 129}$,    
D.K.~Abhayasinghe$^\textrm{\scriptsize 91}$,    
S.H.~Abidi$^\textrm{\scriptsize 164}$,    
O.S.~AbouZeid$^\textrm{\scriptsize 39}$,    
N.L.~Abraham$^\textrm{\scriptsize 153}$,    
H.~Abramowicz$^\textrm{\scriptsize 158}$
% ...
\bigskip
\\

% ...
$^{13}$Institute of Physics, Azerbaijan Academy of Sciences, Baku; Azerbaijan.\\
% ...
$^{34}$$^{(a)}$Facult\'e des Sciences Ain Chock, R\'eseau Universitaire de Physique des Hautes Energies - Universit\'e Hassan II, Casablanca;$^{(b)}$Centre National de l'Energie des Sciences Techniques Nucleaires (CNESTEN), Rabat;$^{(c)}$Facult\'e des Sciences Semlalia, Universit\'e Cadi Ayyad, LPHEA-Marrakech;$^{(d)}$Facult\'e des Sciences, Universit\'e Mohamed Premier and LPTPM, Oujda;$^{(e)}$Facult\'e des sciences, Universit\'e Mohammed V, Rabat; Morocco.\\
% ...
$^{39}$Niels Bohr Institute, University of Copenhagen, Copenhagen; Denmark.\\
% ...
$^{91}$Department of Physics, Royal Holloway University of London, Egham; United Kingdom.\\
% ...
$^{99}$CPPM, Aix-Marseille Universit\'e, CNRS/IN2P3, Marseille; France.\\
% ...
$^{125}$Homer L. Dodge Department of Physics and Astronomy, University of Oklahoma, Norman OK; United States of America.\\
% ...
$^{129}$LAL, Universit\'e Paris-Sud, CNRS/IN2P3, Universit\'e Paris-Saclay, Orsay; France.\\
% ...
$^{153}$Department of Physics and Astronomy, University of Sussex, Brighton; United Kingdom.\\
% ...
$^{158}$Raymond and Beverly Sackler School of Physics and Astronomy, Tel Aviv University, Tel Aviv; Israel.\\
% ...
$^{164}$Department of Physics, University of Toronto, Toronto ON; Canada.\\
% ...
$^{a}$ Also at Borough of Manhattan Community College, City University of New York, NY; United States of America.\\
$^{b}$ Also at California State University, East Bay; United States of America.\\
$^{c}$ Also at Centre for High Performance Computing, CSIR Campus, Rosebank, Cape Town; South Africa.\\
$^{d}$ Also at CERN, Geneva; Switzerland.\\
$^{e}$ Also at CPPM, Aix-Marseille Universit\'e, CNRS/IN2P3, Marseille; France.\\
% ...
$^{*}$ Deceased

\end{flushleft}


\end{document}